%% file: main.tex
\documentclass[a4paper,11pt]{article}
\pdfoutput=1
\usepackage{jcappub}

\usepackage{amsmath}
\usepackage{bm}
\usepackage{xcolor,color,colortbl}
\usepackage[utf8]{inputenc}
\usepackage{hyperref}
\usepackage{url}
\usepackage{graphicx}
\usepackage{multirow,bigdelim}
\usepackage{amssymb}
\usepackage[amssymb]{SIunits}
\usepackage{multirow}
\usepackage{bbold}
\usepackage{verbatim}
\usepackage{comment}
\usepackage{caption}
\usepackage{physics}
\usepackage{subcaption}
\usepackage{orcidlink}

\definecolor{darkgreen}{HTML}{339933}

\usepackage{scalerel,tikz}
\usetikzlibrary{svg.path}
\definecolor{orcidlogocol}{HTML}{A6CE39}
\tikzset{orcidlogo/.pic={
 \fill[orcidlogocol] svg{M256,128c0,70.7-57.3,128-128,128C57.3,256,0,198.7,0,128C0,57.3,57.3,0,128,0C198.7,0,256,57.3,256,128z};
 \fill[white] svg{M86.3,186.2H70.9V79.1h15.4v48.4V186.2z}
 svg{M108.9,79.1h41.6c39.6,0,57,28.3,57,53.6c0,27.5-21.5,53.6-56.8,53.6h-41.8V79.1z M124.3,172.4h24.5c34.9,0,42.9-26.5,42.9-39.7c0-21.5-13.7-39.7-43.7-39.7h-23.7V172.4z}
 svg{M88.7,56.8c0,5.5-4.5,10.1-10.1,10.1c-5.6,0-10.1-4.6-10.1-10.1c0-5.6,4.5-10.1,10.1-10.1C84.2,46.7,88.7,51.3,88.7,56.8z};
}}
\newcommand\orcidicon[1]{\href{https://orcid.org/#1}{\mbox{\scalerel*{
\begin{tikzpicture}[yscale=-1,transform shape]
\pic{orcidlogo};
\end{tikzpicture}
}{|}}}}

\title{Clustering analysis of medium-band selected high-redshift galaxies}

\input{IBIS_author_list}

\emailAdd{ebina@berkeley.edu}

\abstract{
Next-generation large-scale structure spectroscopic surveys will probe cosmology at high redshifts $(2.3 < z < 3.5)$, relying on abundant galaxy tracers such as Ly$\alpha$ emitters (LAEs) and Lyman break galaxies (LBGs). Medium-band photometry has emerged as a potential technique for efficiently selecting these high-redshift galaxies. In this work, we present clustering analysis of medium-band selected galaxies at high redshift, utilizing photometric data from the Intermediate Band Imaging Survey (IBIS) and spectroscopic data from the Dark Energy Spectroscopic Instrument (DESI). We interpret the clustering of such samples using both Halo Occupation Distribution (HOD) modeling and a perturbation theory description of large-scale structure. Our modeling indicates that the current target sample is composed from an overlapping mixture of LAEs and LBGs with emission lines. Despite differences in target selection, we find that the clustering properties are consistent with previous studies, with correlation lengths $r_0\simeq 3-4\,h^{-1}$Mpc and a linear bias of  $b\sim1.8-2.5$. Finally, we discuss the simulation requirements implied by these measurements and demonstrate that the properties of the samples would make them excellent targets to enhance our understanding of the high-$z$ universe. 
}

\begin{document}
\maketitle
\flushbottom

\section{Introduction} \label{sec:introduction}

The large-scale structure (LSS) of the universe can be used as a probe of fundamental physics, cosmology and galaxy formation containing as it does information from inflation, cosmic expansion, gravitational evolution and the relationship between galaxies and the cosmic web. 
Spectroscopic LSS surveys, in particular, precisely map galaxies in the radial direction, allowing for 3D clustering as a function of redshift. This allows observers to collect information of more modes than in 2D and gain access to the dynamics of the universe at late times. 

The current generation of (``Stage IV'') spectroscopic surveys measure 3D galaxy clustering over a significant fraction of the extragalactic sky approaching sample variance precision on quasi-linear scales up to redshift $\lesssim 2$ \cite{DESI,Euclid,PFS}. Measuring millions of galaxy and quasar spectra \cite{DESI-DR1}, these surveys enable a precise measurement of galaxy clustering and e.g.\ sub-percent constraints on the expansion history of the Universe \cite{DESI-DR2}. 
With rapidly improving detector technologies, the precision cosmology frontier can be advanced to even higher redshift, providing improvements in multiple directions simultaneously \cite{Wilson19,Sailer21,Schlegel22,Ferraro22,Beseuner25}. 
DESI-II, a proposed successor experiment to DESI probing galaxies with redshifts up to $\sim3.5$, will be the first step towards a next-generation spectroscopic instrument designed to efficiently map the high redshift Universe \cite{Schlegel22,Bacon24,Beseuner25}.  

Moving beyond $z\sim2$ requires us to trace LSS with new galaxy tracer populations. The LSS probes usually used at $z\lesssim 2$, namely Luminous Red Galaxies (LRGs), Emission Line Galaxies (ELGs; including H$\alpha$ galaxies), and Quasars (QSOs) become sparse past $z\simeq 2$ \cite{DESI,Euclid}.  The number of bright QSOs that can be used to probe the intergalactic medium also declines.  Rather, promising tracers at high redshifts are two sets of galaxies: Ly$\alpha$ emitters (LAE; \cite{Ouchi20}) and Lyman break galaxies (LBG; \cite{Giavalisco02,Shapley11,Ono18,Forster20,Harikane22}). 
These galaxy tracers are defined by Ly$\alpha$ emission and the Lyman break feature in the galaxy spectra (respectively) and have been readily observed to $z\gtrsim 5$ \cite{Ouchi08,Konno16,Malkan17,Sobral18,Ono18}, supplying targets for the full redshift extent of next-generation (``Stage V'') LSS surveys. 

Existing surveys have already demonstrated the feasibility of using these galaxies as cosmological tracers.  Narrow-band imaging with Subaru Suprime-Cam \cite{Ouchi08}, Hyper Suprime-Cam in the SILVERRUSH program \cite{Ouchi18} and, most recently, the One-hundred-square-degree DECam Imaging in Narrowbands (ODIN) survey \cite{Lee24} detects LAEs out to $z\sim7$.  Spectroscopic observations with the VIMOS VLT Deep Survey (VVDS) \cite{LeFevre13}, Hobby-Eberly Telescope Dark Energy Experiment (HETDEX) \cite{Cooper23} and, over the ODIN fields with DESI \cite{DESI,White24} has confirmed thousands of these sources.  For LBGs, wide-field, broadband imaging that employs the ``dropout'' technique -- e.g.~CFHTLS-Archive-Research Survey (CARS) \cite{Hildebrandt09} and Subaru GOLDRUSH \cite{Ono18} -- identifies sources to $z\sim7$, while spectroscopic programs such as VIMOS Ultra-Deep Survey (VUDS) \cite{LeFevre15} and, more recently, DESI \cite{Ruhlmann-Kleider24} have obtained secure redshifts for many of them at $z\lesssim6$.
Although these tracers have been defined through their distinct observational features, they are physically similar objects across a wide range of physics. LAEs and LBGs form a continuum of Ly$\alpha$ rest-frame equivalent width (REW), with LAEs populating the high-REW end of the continuum \cite{Ouchi20,Shapley11}. 

In this work, we report observational results from the DESI ancillary campaign in 2024B, which provided spectroscopic follow-up of medium-band selected high-redshift galaxies. The spectroscopic data provide information on the interloper fraction and redshift distribution of galaxies that can be used to analyze the clustering of the photometrically selected `target' galaxy samples. 
These findings not only establish direct constraints on clustering statistics and offer valuable observational insights, but also represent a key step in developing the forecasting and analytical infrastructure necessary for continued high-redshift investigations. 
This work is done concurrently with related work that together aims to characterize tracers relevant for future high-$z$ surveys. Refs.~\cite{Ruhlmann-Kleider24,Payerne24} have performed investigation of selection and clustering on broad-band selected LBGs, ref.~\cite{Raichoor25} investigates the medium-band selection of high-$z$ galaxies using Subaru Suprime-Cam, and there is ongoing work characterizing the astrophysical properties of LAEs using DESI data.

The present paper is organized as follows. First in \S\ref{sec:tracers} we will describe the high-redshift tracers. In \S\ref{sec:data}, we will briefly describe the photometric and spectroscopic datasets used for the galaxy catalog in this work and use them to characterize where the sample stands with respect to traditional selections. In \S\ref{sec:clustering} we describe the clustering statistics that we use for analysis including an HOD-based analyses of these samples. The mock catalogs populated with representative HODs are then used to study the properties of these tracers (\S\ref{sec:mocks}) and to provide forecasts for future prospects (\S\ref{sec:forecast}). Furthermore, we explore other forms of HODs in \S\ref{app:HMQ} and conclude in \S\ref{sec:conclusion}.  A series of appendices provide supporting details.  Throughout we shall use comoving $h^{-1}$Mpc units, unless otherwise specified, assume a $\Lambda$CDM cosmology using the best-fit values from {\it Planck} \cite{PCP18}, and quote all magnitudes in the AB system  corrected for dust using the SFD98 \cite{SFD} maps.

\section{Tracers at high redshift} \label{sec:tracers}

In this section we describe the galaxy tracers that are populous at high redshifts, of interest for next-generation experiments. The tracers come in two categories: Ly$\alpha$ emitters (LAE \cite{Partridge67}) and Lyman break galaxies (LBG \cite{Guhathakurta90}). Both tracers have been extensively studied previously \cite{Giavalisco02,Ouchi08,Hildebrandt09,Nagamine10,Ouchi10,Shapley11,Garel15,Konno16,Malkan17,Sobral18,Ono18,Gurung-Lopez19,Wilson19,Harikane22}, however in the cosmology community they are relatively unfamiliar due to the expense of obtaining redshifts of faint galaxies over large volumes to pursue classical LSS studies.  Recently, however, photometrically selected LBGs, in combination with CMB lensing, have been used to make a measurement of the growth of structure at high redshift  \cite{Miyatake21}. 

\subsection{Lyman alpha emitters}

LAEs are galaxies characterized by their strong Ly$\alpha$ emission line (at rest-frame 1216\AA) often with low continuum levels, with criteria frequently placed on the minimum rest-frame equivalent width (REW) of the Ly$\alpha$ line. Although this designation is primarily observational, LAEs have been physically identified to be young, low-mass ($M_\star\sim 10^{8-9} M_\odot$), actively star-forming (SFR $\sim1-10\, M_\odot\, \text{yr}^{-1}$) galaxies \cite{Ouchi20}.  Physically LAEs are a subset of the LBG population, however due to the differences in selection and magnitude/mass ranges probed they are often considered to be a separate class.

LAEs are well motivated density tracers for next-generation large-scale structure experiments for multiple reasons. 
First, astronomy surveys have identified populations of LAEs reaching beyond the redshift range of Stage V spectroscopic experiments ($z\lesssim5$) \cite{Umeda25}, with high enough density to enable precise clustering measurements over most of the linear and quasi-linear regimes \cite{Ebina24}.
Second, the spectra of LAEs, with a single emission line on an otherwise featureless spectrum, provide relatively simple targets to select photometrically and obtain spectroscopic redshifts. By selecting for a single big feature in the galaxy spectrum, it is relatively easy to make a simple, yet effective, photometric selection of LAEs, which can limit the effect of systematics in the clustering signal. 
In the astronomy context, most color cuts have been defined using a combination of deep narrow (or medium, although less popular) bands and broad-bands \cite{Sobral18,Umeda25}. These selections primarily focused on obtaining low interloper samples with enough sample size for a statistical analysis of galaxies. Recent work from the ODIN survey \cite{Lee24} has followed this approach, as a low interloper sample is important for angular clustering analyses \cite{White24,Herrera25}. For future spectroscopic cosmology surveys, however, it will be important to cover more volume (i.e.~wider redshift ranges) at the expense of some interlopers, which can be eliminated by spectroscopic follow-up. In addition to simply using medium bands in combination with broad bands, recent work using the Subaru Suprime dataset \cite{Raichoor25} have shown that some selections require only (a closely distributed set of) medium bands.

However, study of LAE in the context of cosmology introduce non-negligible complications as well. One straightforward issue is the non-trivial overlap between traditional LAEs using (deep) narrow bands and LAEs optimal for next-generation cosmology sruveys, using (shallow) medium bands. While both types of selections are designed to target similar-shaped spectra, there has been no direct indication that the different methods are selecting identical galaxies. In particular, this complicates interpretation of past LAE studies.

Another subtlety, perhaps more complicated, is the radiative transfer (RT) of Ly$\alpha$ photons through H{\sc i} in the galaxy environment. Ref.~\cite{Zheng11} suggested that RT can significantly distort target selections depending on local overdensity and LOS velocity divergence, while later work \cite{Behrens18} argued that the magnitude of this effect was largely an artifact of limited simulation resolution. Recent work \cite{Khoraminezhad25} has also shown that RT can significantly suppress LAEs that are central galaxies in massive ($M_h\gtrsim10^{12}M_\odot$) halos.  While studies have not yet measured this RT effect, nor have they settled on a complete model, the impact of H{\sc i} absorption on LAE clustering has been detected \cite{Momose21}.  Simple Fisher forecasts suggest that the impact of a simple RT model can be limited by using a multi-tracer cosmology analysis \cite{Ebina24}. Further investigation is necessary to fully assess the influence of RT on future cosmology surveys. 

\subsection{Lyman break galaxies}

LBGs, also referred to as ``dropout'' galaxies, are a population of galaxies characterized by the Lyman break (rest-frame 912\AA) in the galaxy UV continuum, which arises from the spectral absorption from neutral hydrogen in stellar atmospheres and the interstellar medium \cite{Giavalisco02,Shapley11,Forster20}. For galaxies with redshift $z\gtrsim2$, the Lyman break feature is broadened towards 1216\AA\ due to absorption by neutral Hydrogen in the IGM (i.e.\ `line blanketing' from the Ly$\alpha$ forest) \cite{Shapley11,Wilson19}.
One then searches for a large drop in flux between the `dropout' band (containing $(1+z)\,1000$\AA) and the `detection' band (containing $(1+z)\,1500$\AA) with an almost flat continuum to redder wavelengths.  As with the LAEs, the definition is primarily observational but it effectively selects massive, actively star-forming galaxies at high redshift.
We refer the reader to refs.~\cite{Giavalisco02,Shapley11,Forster20} for a more in-depth and broad study of LBGs in an astronomical context. 

As with the LAEs, there are strong observational reasons to consider LBGs as a central tracer candidate in future high-redshift cosmology experiments. First, LBGs have been confirmed to populate the entire redshift range for potential cosmology experiments in the next decade and beyond, with candidates up to $z\sim9$ discovered with the \textit{James Webb Space Telescope} \cite{Bouwens22}. At the redshifts of interest to next-generation experiments, LBGs will provide sufficient densities to conduct full-shape 3D power spectrum analyses to quasi-linear regimes \cite{Wilson19}. 
LBGs have also been found to have significantly stronger clustering than LAEs, providing a potential opportunity to break degeneracies using the different parameter dependencies (``multi-tracer'' cosmology) \cite{Wilson19,Ebina24}.  For parameters that benefit from sample variance cancellation, such as local-type primordial non-Gaussianity, $f_{NL}^{\rm loc}$, having a high bias tracer and a second tracer with lower bias significantly improves the constraints, by a factor of $\approx 2$ or greater \cite{Sullivan23,Payerne24}.

Traditionally, these galaxies are selected in broadband color spaces using three adjacent broadbands with the ``drop'' in the first two bands and close-to-zero color in the second two bands (with refs.~\cite{Malkan17,Ono18,Ruhlmann-Kleider24,Crenshaw25} providing recent examples). However, using medium-band filters from the Suprime-Cam \cite{Miyazaki02} on the Subaru Telescope \cite{Kaifu00,Iye04}, it has recently been demonstrated that it is possible to select these galaxies in the absence of broadbands \cite{Raichoor25}.  This has partly motivated our study.

\section{Data} \label{sec:data}

In this section we provide an overview of the high redshift galaxy sample that we employ in this work. 
The main drivers are the Intermediate Band Imaging Survey (IBIS) \cite{IBIS} and Dark Energy Spectroscopic Instrument (DESI) \cite{DESI}, providing respectively the bulk of the imaging and spectroscopic data. The imaging data is also supplemented by the public Subaru Hyper-Suprime Cam (HSC) Strategic Survey Program (SSP) \cite{HSC-SSP} dataset. 

\subsection{Imaging data}

The imaging data used for target selections is mainly from the medium-band imaging survey IBIS (co-PI's: A. Dey \& D. Schlegel, NOAO Proposal \# 2023B-184194) on the Dark Energy Camera (DECam) \cite{DECam} at the Blanco telescope in Chile, with assistance from broadband photometry in the HSC-wide survey. 
IBIS is a wide-field, medium-band imaging survey, that is planned to cover $3000\deg^2$ \cite{IBIS}. The observations started in 2024A\footnote{The observations are traditionally described in the `A’ and `B’ semesters, covering January to June and July to December, respectively.}, with the two deep fields (XMM-LSS and COSMOS, each of $\sim$10 deg${}^2$) completed in 2025A. Each deep field is obtained via 92 dithered exposures in each band, achieving homogeneous depth inside a disk of radius $\sim 1.6^\circ$. For this work, we consider the dataset from the XMM-LSS field, which was completed and spectroscopically followed up (\S\ref{sec:DESI}) in 2024B.

\begin{figure}
    \centering
    \includegraphics[scale=1]{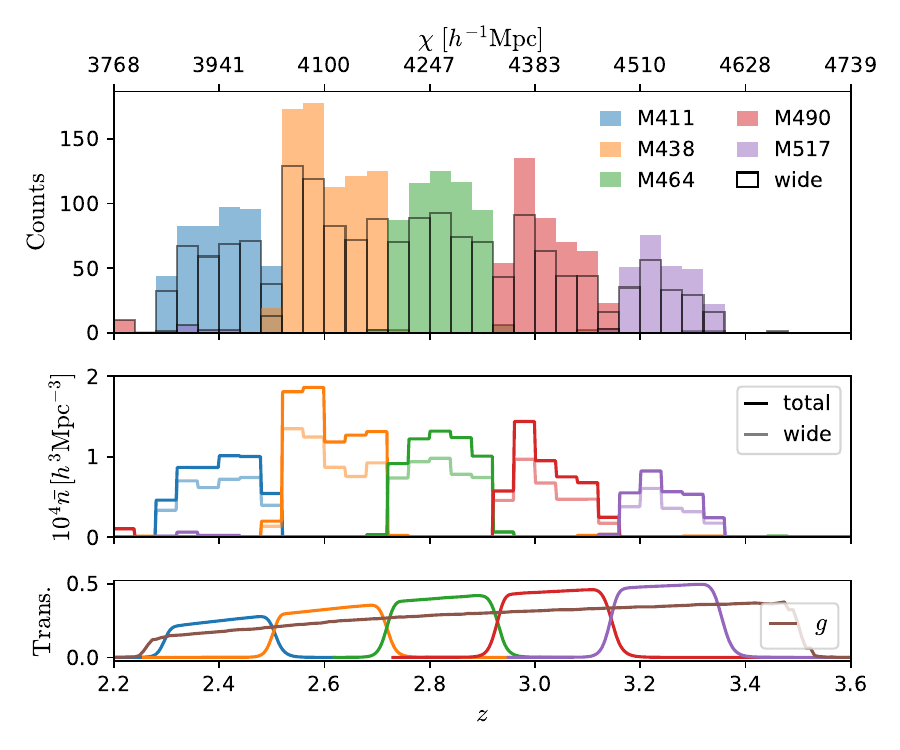}
    \caption{The spectroscopic redshift distribution of the high redshift samples in this work (both the total and wide samples) colored by the medium band used to select it (top), the inferred 3D number density (middle), and the corresponding filter curves (bottom), plotted against redshift assuming $\lambda=(1+z)1216$\AA. The $g$ band is added on the bottom panel for comparison. 
    }
    \label{fig:dNdz}
\end{figure}

As shown in Figure \ref{fig:dNdz}, the filter set of IBIS consists of 5 adjacent medium bands, spanning roughly $4000 <\lambda < 5300$\AA, each with a width of $\sim 260$\AA. In the XMM-LSS field, IBIS provides medium band imaging to a magnitude of 25.5 ($5\,\sigma$ PSF depth). Of the total XMM-LSS field, we restrict the sample to a $1.4\,\deg$ radius to reduce systematics from lower depth in the photometry near the field edges. When accounting for area lost due to masked pixels, bad data or bright object avoidance, the field covers $6.11\, \deg^2$.
We complement these medium-band data with forced-photometry broadband imaging from the HSC-SSP wide field, which reaches depths of $g=26.5$, $r=26.0$, $i=26.5$, and $z=25.5$ \cite{HSC-SSP}.

The data reduction and photometry is performed with the \texttt{legacypipe}\footnote{\url{https://github.com/legacysurvey/legacypipe}} \cite{Lang25} pipeline and the \texttt{Tractor} software\footnote{\url{https://github.com/dstndstn/tractor}} \cite{Lang16}. \texttt{legacypipe} is a post-processing pipeline that performs source detection, wraps \texttt{Tractor} for measurement, and generates catalogs (see \S8 of ref.~\cite{Dey19} for more discussion). For each source the properties, such as position and flux, are obtained by $\chi^2$ minimization of the model images over the relevant region. Tractor produces both ``total'' flux and a ``fiber'' flux for each source, with the latter being the predicted flux within a $1.5^{\prime\prime}$ diameter for $1^{\prime\prime}$ Gaussian seeing.  The fiber fluxes approximate the typical amount of light that a DESI fiber would see, and thus are highly correlated with the SNR.

\subsection{Target selection} \label{sec:selection}

\begin{table}[]
    \centering
    \begin{tabular}{c|c|c|c|c|c|c|c}
        Target sample & $N_{\rm targ}$ & $N_{\rm obs}$ & $N_z$ & $N_{z{\rm good}}$ & $f_{\rm int}$ & $z$ range & $10^2\,\mathcal{L}^{-1}$ \\
        \hline
        M411 & 1065 & 559 & 508 & 457 & 0.10-0.18 & (2.26, 2.56) & 0.55 \\
        M438 & 1938 & 991 & 824 & 736 & 0.11-0.26 & (2.47, 2.77) & 0.61\\
        M464 & 1226 & 631 & 588 & 552 & 0.06-0.13 & (2.68, 2.98) & 0.69\\
        M490 & 891 & 520 & 471 & 436 & 0.07-0.16 & (2.89, 3.20) & 0.78 \\
        M517 & 458 & 284 & 274 & 253 & 0.08-0.11 & (3.10, 3.41) & 0.86 \\
        \hline
        M411 wide & 707 & 395 & 385 & 338 & 0.12-0.14 & (2.26, 2.56) & 0.55  \\
        M438 wide & 1213 & 672 & 588 & 509 & 0.13-0.24 & (2.47, 2.77) & 0.62 \\
        M464 wide & 860 & 459 & 437 & 406 & 0.07-0.12 & (2.68, 2.98) & 0.67 \\
        M490 wide & 606 & 364 & 338 & 303 & 0.10-0.17 & (2.89, 3.20) & 0.77 \\
        M517 wide & 320 & 194 & 192 & 172 & 0.10-0.11 & (3.10, 3.41) & 0.87 \\
    \end{tabular}
    \caption{The observational properties of the high redshift galaxy sample considered for this work. The data are collected over an area of $6.11\deg^2$ in the XMM-LSS field, with counts $N$ being the total number of such galaxies (see text for definitions). Of the entire sample, 54\% of the targets receive sufficient observation time ($\texttt{EFFTIME>2HR}$) and 48\% have secure spectroscopic redshifts ($\texttt{EFFTIME>2HR}\,\&\,\Delta\chi^2>30$), providing information of the redshift distribution and interloper fraction for a `characteristic' subset. $f_{\rm int}$ is the interloper fraction, estimated from DESI spectroscopic follow-up. The range of $f_{\rm int}$ reflects the uncertainty in value from the optimistic and pessimistic estimate over inconclusive redshift evaluations (see main text for details). The $z$-range is determined by the edges of the filter curves. The last column, $\mathcal{L}^{-1}$ in $h\,\mathrm{Mpc}^{-1}$, indicates the impact of projection in diluting the angular clustering, as discussed in \S\ref{sec:clustering}. A narrower redshift bin results a larger multiplier, implying a larger angular clustering given the same underlying distribution of galaxies.}
    \label{tab:data}
\end{table}

\begin{figure}
    \centering
    \includegraphics[width=\textwidth]{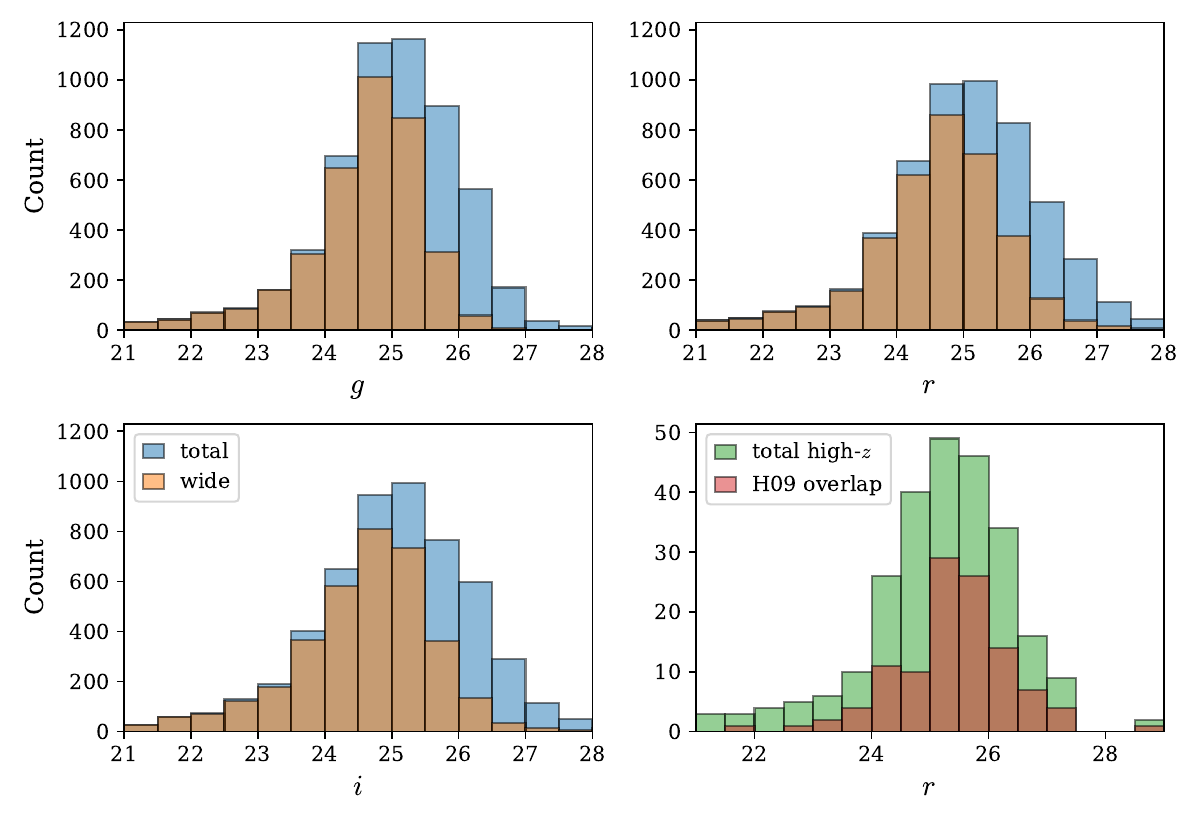}
    \caption{The broadband magnitude distributions of the medium-band selected samples.  We see the $g$, $r$ and $i$-band magnitudes are generally fainter than the MB flux cuts of 25 and 25.3 magnitudes, entirely as expected for samples with emission lines.  In the lower-right panel we compare the $r$-band magnitude distribution of the galaxies that we check for LBG overlap in \S\ref{sec:overlap} (M490 and M517 galaxies that overlap the footprint of CARS \cite{Hildebrandt09}; green) to those that overlap with the LBGs selected by ref.~\cite{Hildebrandt09} (red).}
    \label{fig:broadband}
\end{figure}

As this is the first time we selected galaxies from the IBIS medium bands, the target selection in this work is preliminary, using a combination of several novel techniques using the consecutive configurations of medium bands. The goal is to have a selection optimized towards the next wide-field spectroscopic survey, likely DESI-{\sc ii}, with the current selections aiming towards exploring the parameter space rather than a final tuning for efficiency. Based on ongoing investigation, we believe that we can increase the size of the target sample through improved selection techniques. In the meantime, an analysis of these objects provides important information for future planning.

The target selection in this work is a combination of a number of different, exploratory selections. While we defer details about the selections to companion papers \cite{Raichoor25,Dey25}, we will describe them briefly here. We used three of the selections explored to date, each of which can be performed over multiple medium bands. The first category relies on the colors between three adjacent medium bands to select a slight peak in the center band, e.g.\ to select objects with an emission line in the M464 filter
\begin{equation}
    \text{M438}-\text{M464}>0.6\  , \ 
    \text{M464}-\text{M490}<0\ ,  \  
    \text{M438}-\text{M464}> 1 + 0.6 (\text{M464}-\text{M490})
\end{equation}
where M438, M464, and M490 are the magnitudes in their respective filters (Fig.~\ref{fig:dNdz}).  An additional `quality' cut (\texttt{FRACFLUX}$<0.15$) is applied for each medium band to reject objects with significant flux contributions from adjacent sources. 
The second category performs a traditional LAE color cut by using the five medium bands to construct a synthetic $g$ band and placing a cut on the excess color between a medium band and the synthetic broad band, e.g.
\begin{equation}
    \text{M464}-g_\text{synth} < -0.5 \quad , \quad
    \text{M464}-g_\text{synth} < 0.5 - 2.8 \exp{\text{M464}_\text{fib}-26}
\end{equation}
where $\text{M464}_{\rm fib}$ is the fiber magnitude in the M464 filter.  
The third category, similar to the second, uses the color between an individual medium band and a synthetic band, averaging the flux of the other four medium bands, e.g.
\begin{equation}
    \text{M464}-\text{synthetic}<-0.45 \ , \ 
    \text{M464}-\text{synthetic}<-0.5 - 0.7 (\text{M464}-24.2)
\end{equation}
We combine these three selections to construct the target sample used in this work. 

By targeting the Ly$\alpha$ emission line at $1216$\AA, each of these selections allow for an efficient selection of high-$z$ galaxies in bins of $\Delta z\sim0.2$. Together, they form a catalog spanning $2.26<z<3.41$.
In this work we consider a subset\footnote{There is a non-negligible fraction ($22\%$) of our targets that were not part of the original target sample. This is due to a coding error in one of the initial target selections, which has been corrected by substitution with a similar, but nonidentical selection.} of the initial target sample used for spectroscopic follow-up, in order to reduce and better constrain interloper fractions, which is important for the angular clustering that we will perform. 
For the latter three medium bands (that we analyze below as the $z=3$ bin) we have applied tighter fiber magnitude cuts to each selection (the faintest at 25.3), as well as additional broadband color cuts using the $g$, $r$ and $i$ bands
\begin{equation}
    g-r<0 \quad | \quad (r-i<0.3 \quad \& \quad g-r<0.5)
    \quad .
\end{equation}
At the expense of reducing the sample size by a factor $\sim2$, this effectively distinguishes true high-$z$ galaxies from interlopers in the stellar locus, reducing the interloper fraction from $f_{\rm int}\lesssim0.4$ to $\lesssim0.15$ and reducing the uncertainty in the interloper fraction by a factor of $\sim3$. For the first two medium bands (that we will analyze as the $z=2.5$ bin), the interloper fractions are significantly lower ($f_{\rm int}\lesssim0.25$) than the higher redshift medium bands, and hence we do not perform refinement in order to preserve the statistical power of the sample\footnote{When performing a high-purity refinement for the first two medium bands, we found that the detection of clustering was $\lesssim1\sigma$ due to the large increase in shot noise.}.

The properties of our samples are given in Table \ref{tab:data}.
For each filter we define two subsets of the full sample, with the `total' sample reaching a medium-band fiber magnitude of 25.5 ($z=2.5$ bin) or 25.3 ($z=3$ bin), and the `wide' sample to $25$.  The `wide' sample will provide a direct reference for the ``wide-field'' component of the IBIS survey, which will observe to medium-band depths of $25$. Note that the intention here is not to simulate the target selection for the wide sample, as that will require degrading of the medium-band signals, but to provide reference to the luminosity limit of the brighter wide field. 
In Figure \ref{fig:broadband} we show the distribution of broadband magnitudes for the galaxy samples. Notice that they are considerably fainter than suggested by the medium-band magnitude limits, extending beyond $r\sim26$. This is as expected for samples that contain significant Ly$\alpha$ line emission.

\subsection{DESI spectroscopy} \label{sec:DESI}

\begin{figure}
    \centering
    \includegraphics[width=.97\linewidth]{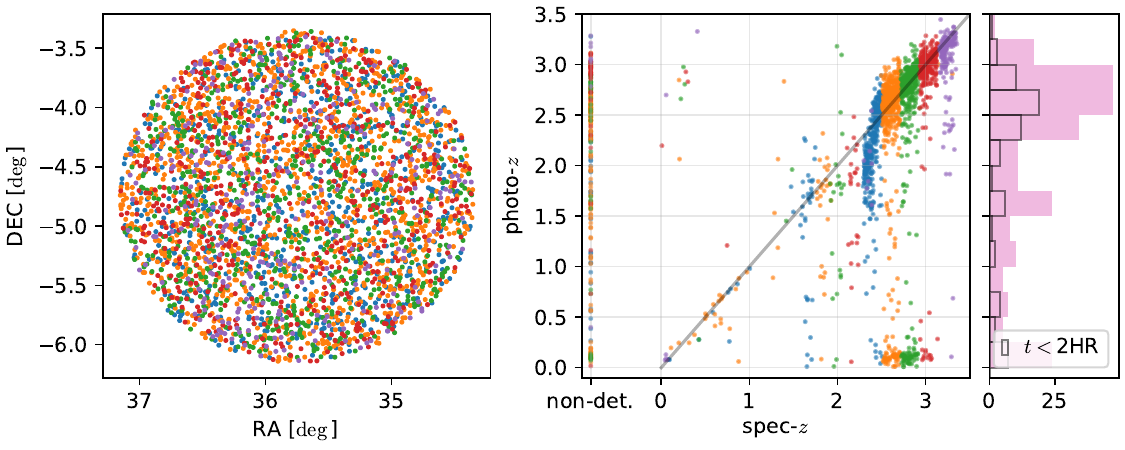}
    \caption{Left: The distribution of selected targets on the sky, with colors corresponding to the filters in Figure \ref{fig:dNdz}. Center: The performance comparison between CLAUDS SExtractor LePhare photometric redshifts, which were used to inform the target selection, and the DESI spectroscopic redshifts ($\Delta\chi^2>30$ in template fitting software \texttt{RedRock} \cite{Redrock}). Targets with \texttt{EFFTIME>1HR}, but no spec-$z$, are shown on the left end of the figure. The targets are again colored according to their selection bands. Right: The photo-$z$ distribution of objects with observation time $\texttt{EFFTIME}>\texttt{1HR}$, but no reliable spec-$z$. We isolate the objects that do not reach our fiducial \texttt{EFFTIME} in black outlines and find that the duration of observation does not have a qualitative impact on the photo-$z$ distribution of failures.}
    \label{fig:photoz}
\end{figure}

DESI is a Stage-IV cosmology survey designed to study dark energy \cite{Levi13,DESI}. It employs a highly multiplexed spectroscopic instrument stationed on the 4-m Nicholas U.~Mayall Telescope at the Kitt Peak National Observatory in Arizona, USA. The main survey started in 2021 and has collected over 40 million extragalactic redshifts to date. Using 5000 robotic fibers covering a $3.2\,\deg$ diameter field of view, it is able to collect $10^{4-5}$ spectra per night \cite{DESIb,DESI22,Silber23,Miller23,Schlafly23,Poppett24}. Aside from the main survey, DESI allocates time for ancillary programs to pursue a wide range of science goals with (relatively) cheaper fiber cost. We use the ancillary program \texttt{tertiary44} ({\sc TileID}s=83519--83548) to follow-up the targets selected using IBIS medium bands. Here, we describe the details of this observation.

Observations are spread over a series of tiles, each assigning targets to the $5000$ fibers (of which just over $4000$ are for science targets).  Targets are assigned to fibers using information about the targets' positions, the current state of the focal plane and a set of priorities and required number of observations. The observations are `dithered' so that each target obtains data from a number of different fibers to obtain the desired exposure time.

In order to maximize the observation efficiency in the ancillary program, the fiber assignment \cite{Fiberassignment} is done over two iterations -- a pre-assignment and final assignment. The pre-assignment runs a typical fiber assignment algorithm over the proposed targets and determines the targets that are likely to receive the required number of assigned fibers (for this sample, $\texttt{NASSIGN}=8$, which translates to $\texttt{EFFTIME}\simeq\texttt{2HR}$ where \texttt{EFFTIME} is a measure of exposure time accounting for observational conditions -- see \S4.14 of \cite{Guy23}). The targets that are deemed unlikely to reach the required exposure are demoted from the target class in order to assign their allocated fibers to other ancillary targets. While this maximizes the fiber usage efficiency, it poses significant difficulty in terms of the clustering analysis, as discussed further in Appendices \ref{app:scales} and \ref{app:fiber}.

The spectroscopic observations, reduced through DESI's extensive pipeline \cite{Guy23} version \texttt{loa}, provide template-based redshift measurements of the targets using a modified version of \texttt{RedRock} \cite{Redrock} including four high-redshift templates (courtesy of J.~Moustakas as described in \cite{Raichoor25}). While the pipeline’s performance has been validated on earlier datasets \cite{Raichoor25}, we did not carry out a dedicated visual-inspection (VI) campaign for the present sample, so further refinements remain possible. 
The redshift information allows us to determine the three-dimensional positions and quantify key information such as redshift distribution and interloper fractions, as shown in Figure \ref{fig:dNdz} and Table \ref{tab:data}. Here, we define the interlopers to be galaxies that do not lie in the redshift range of the respective medium-band filter (i.e.~an M411 galaxy with $z=2.6$ is an interloper). As we do not know the redshifts of galaxies with inconclusive redshift evaluations, we provide a range of interloper fraction ($f_{\rm int}$) estimates. The optimistic estimate will be attributing these galaxies to the intended MB redshift range ($f_{\rm int}=N_{z \rm good}/N_z$), while the pessimistic estimate will be attributing them to interlopers ($f_{\rm int} = N_{z \rm good}/N_{\rm obs}$). Note that even without the spectroscopic data, we are able to locate most galaxy redshifts to $\Delta z\simeq 0.2$ due to the medium-band selections targeting the Ly$\alpha$ emission line.

Figure \ref{fig:photoz} shows the on-sky positions of the objects making up our total sample, colored by the medium-band filter in which Ly$\alpha$ would lie.  We also compare the photo-$z$ from CLAUDS SExtractor LePhare \cite{Desprez23}, which were originally used in planning the target selection, and the DESI spectroscopic redshift where we have a secure\footnote{Note that the DESI pipeline has not been optimized for redshifting galaxies such as these in this $z$-range so we anticipate that further work could increase the number of secure redshifts non-trivially.} redshift.  Overall the photo-$z$ and spec-$z$ track each other quite well.  The spectroscopic redshifts are concentrated in the ranges defined by the filter bands, as expected: we are more likely to get a secure spec-$z$ when we have Ly$\alpha$ emission in the band.  The distribution of photo-$z$s for the redshift failures suggests that many of them also lie at $z>2$, but this is partly due to the fact that we used the photo-$z$s in designing our targeting cuts.  The number of objects with $z_{\rm phot}>2$ with secure spectroscopic redshifts $z<2$ is quite small ($1\%$), illustrating the effectiveness of the medium-bands for selecting high-redshift targets.

Later we will measure the angular clustering of our targets, and use the $dN/dz$ and $f_{\rm int}$'s determined from spectroscopy to infer the underlying 3D clustering. Of our entire target sample, 48\% of objects have secure ($\texttt{EFFTIME>2HR}\,\&\,\Delta\chi^2>30$) spectroscopic redshifts and hence will be able to provide reliable characterization of its redshift distribution. As we will see, restricting the sample to only objects for which we have a well characterized $f_{\rm int}$ means that the remaining uncertainties on $dN/dz$ and $f_{\rm int}$ are much smaller than our statistical uncertainties.  We shall henceforth neglect them.

\subsection{Overlap with traditional tracers} \label{sec:overlap}

As our selections are dissimilar to many previous high redshift target selections, we would like to understand the composition of the galaxy sample in this work viewed with respect to traditional descriptions of LAEs and LBGs.  Note at the outset that our investigation must be preliminary as the data are prone to significant selection effects and incompleteness with pipelines that are still under development.

All three color cuts described in \S\ref{sec:selection} are designed to pick out Ly$\alpha$ emission, so the resulting galaxies should, on average, exhibit large rest-frame equivalent widths (REWs) and qualify as LAEs under the definition of various literature \cite{Ouchi08,Ouchi10,Gronwall07,Garel15}. Even so, the catalog is not expected to consist of only LAEs, as LAEs and LBGs populate an overlapping spectrum of Ly$\alpha$ emission strength \cite{Shapley11,Ouchi20}. 
To place this discussion into context in the previous literature, we cross-match with the $u$-dropout LBG sample from ref.~\cite{Hildebrandt09} (called the H09 catalog from here on) to find the overlapping objects. For an accurate assessment of the overlapping galaxies, we restrict our cross-matching catalog to the M$490$ and M$517$ samples ($2.89<z<3.41$), which match the narrower redshift distribution ($2.9\lesssim z\lesssim 3.6$) of the H09 sample\footnote{The mismatch in the redshift range $3.41<z\lesssim3.6$ does not concern us, as we are only interested in the overlap of our sample with the H09 catalog.}. 

The cross-matching reveals that 46.1\% of our M490 and M517 galaxies are LBGs, implying that a significant fraction of LBGs have Ly$\alpha$ emissions strong enough to be picked up by the medium bands used in the target selections. This, taking into account the sampling bias from our target selection, aligns with past study of LBGs, that LBG REWs are distributed approximately evenly between positive and negative Ly$\alpha$ emissions, with 25\% having REW$>20\AA$ \cite{Shapley03,Shapley11}. 
The broadband magnitude distribution of the overlapping galaxies, shown in the lower-right panel of Figure \ref{fig:broadband}, reveals that the LBGs present in our catalog follows the distribution of the overall sample and peaks at a relatively faint continuum magnitude of $r\sim25.5$. This can be understood as a consequence of the target selections detecting Ly$\alpha$ peaks through a color excess between Ly$\alpha$ band and adjacent bands. 
While this result sheds light on the mixture of LAEs and LBGs in the present work, extrapolating this result to the total galaxy sample requires care as incorporating the redshift dependence of each sample would be non-trivial. 

\section{Clustering analysis} \label{sec:clustering}

\subsection{Angular clustering} \label{sec:angularclustering}

Although we obtain spectra for a significant fraction of our galaxy samples, we will not be using 3D clustering due to fiber assignment complications, as described in Appendix \ref{app:fiber}. Instead, we will use the target sample to perform a pseudo-3D clustering through layers of angular clustering.  This approach also has the advantage that it can be directly applied to the ``wide'' layer of IBIS without requiring extensive spectroscopic follow-up.

\begin{figure}
    \centering
    \includegraphics[width=0.98\linewidth]{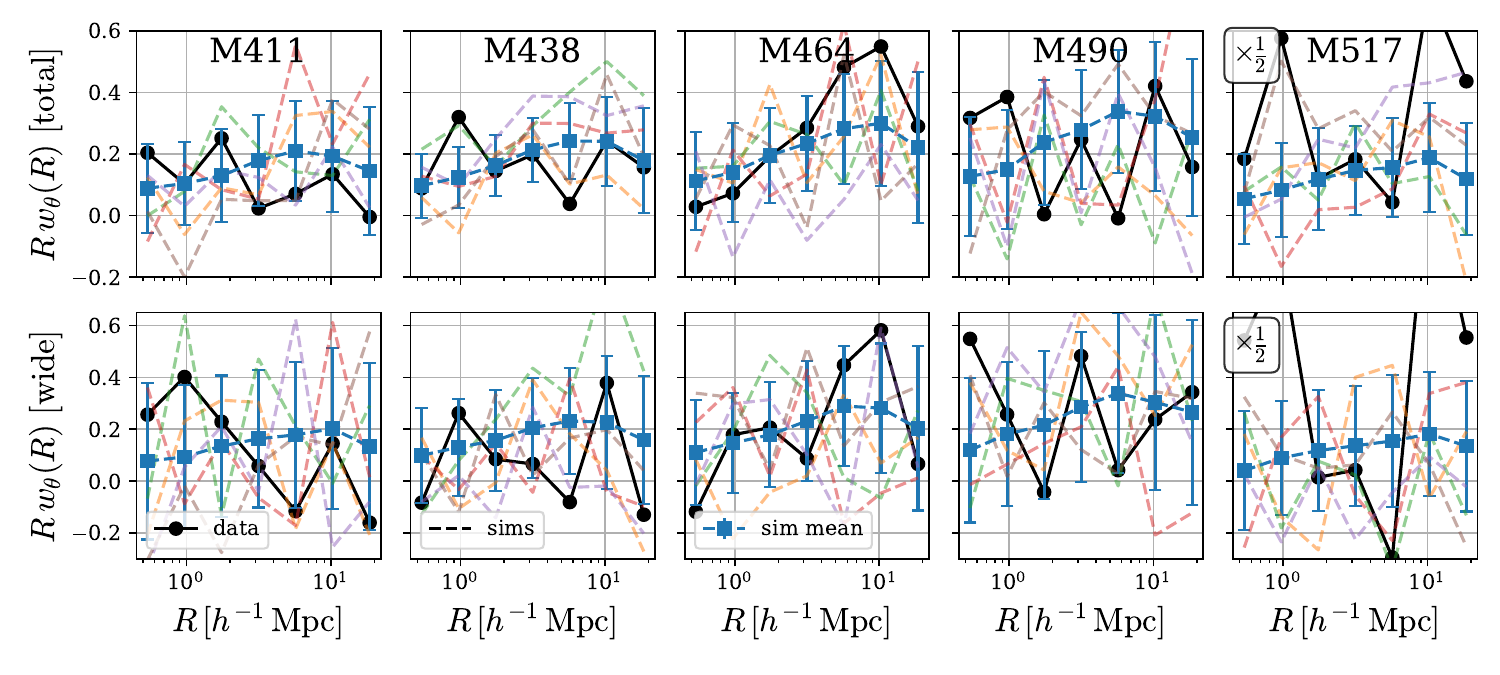}
    \caption{The angular clustering measurements $w_\theta(R)=w(\theta=R/\chi_0)$ of each medium band, with the total and wide sample shown on the top and bottom panels, respectively. The data measurements, in black circles, are shown together with the predictions from the best-fit HOD model (blue squares) and 5 random realizations of the best-fit HOD (faint dashed lines). Note that these measurements and errorbars are not used to directly infer goodness-of-fit to the mock and determine the best-fit HOD. That is done using $w_p(R)$, shown in Figure \ref{fig:wR}.
    The errors are calculated using 256 pseudo-independent realizations of the HOD, by offsetting the observer and observing the simulation through the survey mask and redshift selection function.  For display purposes the black points have been scaled by $1/2$ in the final column, since they exhibit large scatter.
    }
    \label{fig:wt}
\end{figure}

As our target samples naturally come subdivided into slices of width $\Delta z\sim0.2$ ($\Delta \chi\sim200 \,h^{-1}{\rm Mpc}$), we can perform pseudo-3D clustering by collecting the clustering information in each redshift bin.  Specifically we compute the angular auto-correlation function, $w(\theta)$, of the targets within each filter and then combine the resulting measurements into broader redshift bins.  We choose to combine the 5 filters into two $z$ bins at $z\sim2.5$ (M411 and M438) and $z\sim3$ (M464, M490, and M517).
Each $w(\theta)$ corresponds to taking the (distribution-weighted\footnote{The impact of lensing is expected to be highly subdominant to our observational errors.}) radial integral over the 3D correlation function $\xi$
\begin{align}
    w(\theta) &= \int d\chi_1 d\chi_2\ f(\chi_1) f(\chi_2)\, \xi\left(\sqrt{\chi_1^2 + \chi_2^2 - 2\chi_1\chi_2\cos{\theta}} \right) \label{eqn:wtheta-projection} \\
    &\approx \left[ \int d\chi\ f^2(\chi) \right] \int d\Delta\ \xi\left(\sqrt{\chi_0^2 \widetilde{\omega}^2 + \Delta^2} \right)
    = \mathcal{L}^{-1} \ w_p(R=\chi_0\widetilde{\omega})
    \label{wtheta}
\end{align}
where $\widetilde{\omega}=2\sin(\theta/2)\simeq \theta$, $\Delta=\chi_2-\chi_1$ and $f(\chi)$ is the radial (redshift) distribution normalized such that $\int f(\chi) d\chi=1$.  We have defined $\mathcal{L}^{-1}=\int d\chi\, f^2(\chi)$ and note that for a tophat $f(\chi)$ of full-width $\Delta\chi$ we have $\mathcal{L}=\Delta\chi$.  The $\mathcal{L}$ values for each redshift slice are listed in Table \ref{tab:data}. In the second line above we have used that our filters select narrow $f$ peaked around comoving distance $\chi_0\gg \Delta\chi$ and converted from angle to transverse separation using the comoving distance to each slice center (see Appendix \ref{app:medium_shell} or refs \cite{Padmanabhan09,Myers09}).  We recognize the integral of $\xi$ over $d\Delta$ as $w_p(R)$ evaluated at $R=\chi_0\theta$ and with the limits of the line-of-sight integration defined by the width of the redshift slice (i.e.\ approximately $\pm 100\,h^{-1}$Mpc) rather than a fixed maximum separation.  Note that this projection justifies the use of the real-space correlation function in Eq.~\ref{eqn:wtheta-projection}.  We discuss this approximation further, and compare it to the full calculation, in Appendix \ref{app:medium_shell}.

We compute $w(\theta)$ using the Landy-Szalay estimator \cite{LS} in each bin
\begin{equation}
    w(\theta) = \frac{DD-2DR-RR}{RR}
\end{equation}
where $DD$, $DR$ and $RR$ represent (normalized) counts of pairs of data and randoms, with the random points that define the survey footprint and masks generated using the \texttt{desitarget} pipeline within DESI \cite{Myers23}. 

In order to combine information from five medium-band redshift slices into two $z$-bins, we will bin the correlation function in transverse distances rather than angles ($w_\theta(R) = w(\theta=R/\chi_0)$) to estimate the projected correlation function $w_p^{(i)}$ in each bin (Eqn.~\ref{wtheta}) and produce their inverse variance-weighted sum
\begin{align}
    w_p(R) &=  C\sum_i C^{-1}_i \left[ \mathcal{L}^{(i)} w_\theta^{(i)}(R) \right]
\end{align}
where the summation index $i$ runs over the combined redshift slices, $\mathcal{L}^{(i)}\,w_\theta^{(i)}(R)$ and $C_i$ are the projected correlation function and covariance matrix for slice $i$, and $C^{-1}=\sum_i C^{-1}_i$ is the inverse covariance of $w_p$. The $\mathcal{L}$ applied in the calculation of $w_p^{(i)}(R)$ are listed in Table \ref{tab:data}.

\begin{figure}
    \centering
    \includegraphics[width=0.98\linewidth]{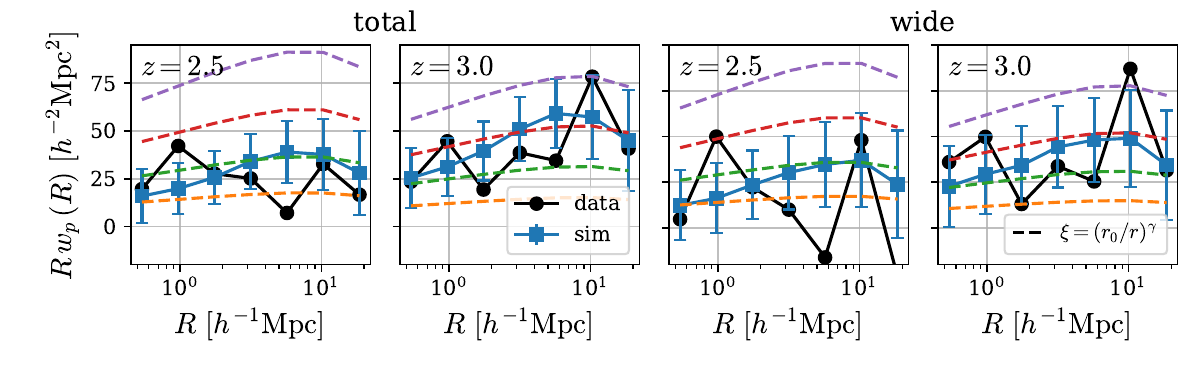}
    \caption{The pseudo-projected clustering, $w_p(R)$, for the $z=2.5$ and 3.0 $z$-bins for each clustering sample. The data (black circles) are compared against the best-fit HOD model (blue, Table \ref{tab:HOD_params}) yielding fits within $\approx1\sigma$. We also overlay the predictions from a power-law correlation function $\xi=(r_0/r)^\gamma$, where the orange, green, red, and purple lines correspond to $r_0 = 2,$ 3, 4, and 5$\,h^{-1}\,\text{Mpc}$. The errors are calculated using 256 pseudo-independent realizations, as in Figure \ref{fig:wt}.}
    \label{fig:wR}
\end{figure}

Although this is not a pure 3D clustering measurement, it includes information about every galaxy pair within each redshift slice and to the extent that these slices are much larger than the clustering distances that we are concerned with ($\Delta\chi \gg r\sim\mathcal{O}(10) \,h^{-1}{\rm Mpc}$) it retains most of the information present in a real-space 3D correlation function.  By integrating in the line-of-sight direction we avoid the need to model redshift-space distortions and by limiting the line-of-sight integration to the width of the medium band we avoid excessive signal loss due to projection.  Importantly, by using the target sample directly we avoid needing to accurately model the impacts of fiber assignment.
The angular clustering for each redshift bin $w_\theta(R)$ are shown in Figure \ref{fig:wt}, along with clustering from the best-fit mock catalog discussed in \S\ref{sec:HOD}. The combined measurements $w_p$, for both (large) redshift bins are shown in Figure \ref{fig:wR}, again along with the mock catalog fit.  For both figures, we have corrected the `raw' measurements up by a factor of $(1-f_{\rm int})^{-2}$ to account for interlopers diluting the signal, under the (reasonable) assumption that the interlopers are spatially uncorrelated with the signal and of lesser clustering amplitude\footnote{Due to this correction, to linear theory order, an error in $f_{\rm int}$ translates into an error in linear bias $b$.}. We use the middle of the $f_{\rm int}$ range quoted in Table \ref{tab:data} for each sample to minimize (the already negligible) uncertainty\footnote{The largest uncertainty in $f_{\rm int}$ is 0.15 for M438 with deep imaging depth. With selecting the middle of the estimated range, the expected error is 0.15/2, which corresponds to $\sim15\%$ amplitude error.}. 

As is apparent from Figure \ref{fig:wR}, we have detected clustering in both $z=3$ samples and the total sample at $z=2.5$, but the detection is marginal for the wide sample at $z=2.5$.  However even this marginal detection is of interest because it provides an upper limit to the clustering amplitude and hence provides information about the host halos of these galaxies that is valuable for planning purposes.  To get a feel for the implied clustering, Figure \ref{fig:wR} shows $w_p$ for four different correlation lengths ($r_0=2$, 3, 4 and $5\,h^{-1}$Mpc) assuming the real-space correlation function $\xi=(r_0/r)^{\gamma}$ with fixed slope $\gamma=1.8$.
The array of curves fit the data successfully relative to our generous errors, indicating that we already have a basic understanding of the galaxies at hand\footnote{Note that while we do not depend explicitly on the best-fit model here, we adopt the covariance matrices of the best-fit HOD models to provide the statistical comparison to data.} and suggest $r_0\approx 2-4\,h^{-1}$Mpc.  We will refine this initial estimate using N-body simulations next, but we note that this is in line with previous LAE studies despite differences in sample definition \cite{White24,Khostovan19,Bielby16,Ouchi10,Gawiser07}. The comparison with previous work on LBGs is less trivial as the correlation length is a clear function of continuum magnitude $m_{UV}$. With this in mind, our rough inference agrees with previous LBG samples of faint ($m_{UV}<25$, 25.5) continuum magnitudes \cite{Hildebrandt09,Bielby11,Durkalec15}.

\subsection{HOD fit} \label{sec:HOD}

In this section, we will use mock catalogs from Halo Occupation Distribution (HOD) models to compute covariance matrices and interpret our clustering measurements. A HOD model assigns (mock) galaxies to dark matter halos through a probability distribution function that depends (primarily) upon halo mass.  The form of the halo occupation for galaxies such as those in our sample is not well understood, nor are there strong observational constraints on the dependence upon secondary properties of halos, beyond mass.  On the other hand, our data are currently unable to provide precise constraints over a wide range of scales and thus many of our inferences will be robust to model (\S\ref{app:HMQ}) and cosmology misspecification.

In this work, we will primarily use the functional form introduced by ref.~\cite{Zheng07} as implemented in the AbacusUtils\footnote{\hyperref[https://abacusutils.readthedocs.io/en/latest/]{https://abacusutils.readthedocs.io/en/latest/}} software \cite{Yuan22} accompanying the \textsc{AbacusSummit} N-body suite \cite{Maksimova21}, which have been generated using the \textsc{Abacus} N-body software \cite{Garrison21}.  We perform comparison to a different HOD form in \S\ref{app:HMQ}.  For each halo, the `standard' model assigns central galaxies with a binomial distribution and a Poisson-distributed number of satellites, with means
\begin{align}
\langle N_{\text{cen}}(M) \rangle &= \frac{1}{2}  \text{erfc} \left( \frac{\ln{M_{\text{cut}}/ M}}{\sqrt{2}\sigma} \right)
\label{eqn:hod-cen} \\
\langle N_{\text{sat}}(M) \rangle &= \langle N_{\text{cen}}(M) \rangle \left( \frac{M - \kappa M_{\text{cut}}}{M_1} \right)^{\alpha}
\quad \text{for } M > \kappa M_{\text{cut}}
\label{eqn:hod-sat}
\end{align}
where $M$ is the halo mass and \{$M_{\rm cut}$, $M_1$, $\sigma$, $\kappa$, $\alpha$\} are model parameters. 
Although this functional form, and HOD models more generally, were originally designed to model magnitude limited samples at $z\lesssim 0.5$, due to the level of flexibility of the model we expect moderate success even for high-redshift galaxies. Indeed, N-body studies and observations have each shown that the halo distribution and stellar mass/halo mass relations are roughly consistent across a wide range of redshifts, indicating that the functional form of the HOD model will be at least roughly applicable at $z\sim3$ \cite{Wechsler18}. Works modeling the high-redshift universe with semi-analytic models or hydrodynamical simulations have also shown that these HOD models are moderately successful at describing high-$z$ galaxies \cite{Nagamine10,Garel15,Gurung-Lopez19,Ravi24,Sullivan25,Khoraminezhad25}.  
While this clearly does not guarantee that these models will correctly mimic the galaxy samples at hand, at the precision level of this study we deem it sufficient. It will be valuable to follow this up in the future with more constraining data, as detailed simulation-based study of medium-band selected high-$z$ galaxies will provide important insight into both future cosmology survey planning and the evolution of galaxies within large-scale structure. 

Within the standard cosmological model, the amplitude of clustering we measure indicates that our galaxies are hosted by halos only slightly more massive than the characteristic mass in the halo mass function.  Since at these redshifts the characteristic mass is quite low, we will construct mock catalogs using the high-resolution simulation of the \textsc{AbacusSummit} suite.  This simulation assumes a Planck $\Lambda$CDM cosmology \cite{PlanckLegacy18,PCP18} and uses $6300^3$ particles in a $1\, h^{-1}{\rm Gpc}$ (periodic) box. This implies a particle resolution of $m_{\rm part}=3.5\times 10^{8}\,h^{-1}\,M_\odot$ \cite{Maksimova21}, which we will see later provides just sufficient resolution for the halos of interest. Given the wavelength coverage of our medium-bands, we will use the $z=2.5$ output to analyze the combination of the first two filters and the $z=3$ output for the combination of the latter three. 

The measurement errors can be estimated by observing 256 pseudo-independent realizations of each model by randomly offsetting the simulation along each axes of the box and ``observing'' them by applying 2D masks and redshift distributions. The masks are constructed from the random catalog using fine (\texttt{nside}=8192) \texttt{healpy} pixels \cite{healpix} -- any pixel that has a random in it is considered observable. The redshift distributions are inferred from the data obtained through DESI spectroscopic follow-up (see Figure \ref{fig:dNdz}). As the observation volume is significantly smaller than the simulation volume (by a factor of $\sim80$ and $\sim60$ for the $z=2.5$ and 3.0 bins, respectively), each random offset provides nearly independent realizations of the density field\footnote{While the number of mock observations is significantly larger than the ratio of simulation volume to survey volume, for the covariance estimation purposes it is unnecessary to have fully independent realizations.}.

\begin{table}[]
    \centering
    \begin{tabular}{c||c||c|c|c|c|c|c|c|c}
        $z$ & Sample & $\log M_{\rm cut}$ &  $\log M_1$ &  $\sigma$ &  $\kappa$ &  $\alpha$ & $\chi^2$ ($n\sigma$) & $f_{\rm sat}$ & $r_0$\\
        \hline
        \hline
        \multirow{2}{1.5em}{2.5} & Total & 11.25 & 11.95 & 0.66 & 1.00 & 0.33 &  7.36 ($0.86\sigma$) &  0.12 & $3.28 \pm 0.70$ \\
        & Wide & 11.00 & 12.00 & 0.66 & 1.00 & 0.50 & 12.10 ($1.66\sigma$) & 0.03 & $3.08 \pm 0.80$ \\ 
        \hline
        \multirow{2}{1.5em}{3.0} & Total & 12.00 & 12.70 & 0.66 & 1.00 & 0.50 & 7.22 ($0.83\sigma$) & 0.10 & $4.13\pm0.24$ \\
        & Wide & 11.75 & 12.45 & 0.66 & 1.00 & 0.33 & 6.65 ($0.73\sigma$) & 0.05 & $3.85\pm0.68$ \\ 
    \end{tabular}
    \caption{The HOD parameters for the best-fit projected correlation functions against data in Figure \ref{fig:wR}, along with the goodness-of-fit $\chi^2$, satellite fraction $f_{\rm sat}$, and correlation length $r_0$ of the best-fit model. $f_{\rm sat }$ and $r_0$ are parameters derived from simulation (see text). The uncertainties in $r_0$ are derived from simulations satisfying $\Delta\chi^{2}<2$. Due to the limited mass resolution, we quote the error corresponding to the upper end of the $r_0$ range (Figure~\ref{fig:r0}). Masses are in $h^{-1}M_\odot$ and distances in $h^{-1}$Mpc.} 
    \label{tab:HOD_params}
\end{table}

Given an estimate of the covariance matrices, we can test any given HOD model for consistency with the data.  Since the data at present have limited constraining power we have not attempted a formal fit.  Instead we scan over a range of HOD parameters\footnote{Specifically we fix $\kappa=1$ and choose $\sigma=\{0.33,0.5,0.66\}$, $\alpha=\{0.33,0.50,0.66\}$ while sampling $\log_{10}M_{\rm cut}/(h^{-1}M_\odot)$ in steps of $0.25$ between $11.0$ and $12.5$ and $M_1/M_{\rm cut}$ in $\{5,10\}$ inspired by the HOD parameters explored in ref.~\cite{White24}. }, testing each to find the best-fit model (in our set) and the range of models that provide acceptable fits.
This allows us to constrain both clustering and galaxy properties. 
The HOD parameters for the best-fit mock catalog from our set are shown in Table \ref{tab:HOD_params}, along with associated goodness-of-fit values, satellite fractions $f_{\rm sat}$, and correlation length $r_0$. The corresponding clustering predictions are shown in Figure \ref{fig:wt} against the data, with five random realizations of the mock catalog to give a rough intuition for individual fluctuations from sample variance. We observe that the highest redshift bin (M517) has the poorest fits in each sample. The reason for this is not clear, but we believe that it owes to the significantly smaller sample size of those bins compared to other bins, as shown in Table \ref{tab:data}.

\begin{figure}
    \centering
    \includegraphics[width=0.97\linewidth]{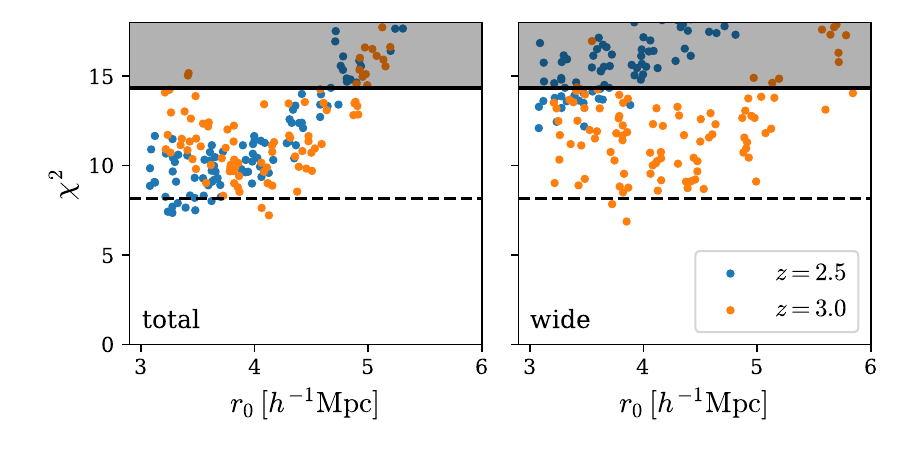}
    \caption{The $\chi^2$ distribution of HOD models plotted against predicted correlation length, $r_0$, with the total and wide samples shown in the left and right panels, respectively. The $1\,\sigma$ and $2\,\sigma$ CL for 7 d.o.f. are shown in dashed and solid lines, respectively, with the region beyond $2\,\sigma$ shaded. The distribution follows a parabolic curve with moderate levels of scatter.  Lower scatter indicates that $r_0$ serves as a good summary statistic for the clustering, containing most of the information (for models following these HOD forms at the current level of precision). That the $z=2.5$ points show no clear minimum in the $\chi^2$ vs.\ $r_0$ is a combination of limited numerical resolution (preventing us from examining lower $r_0$) and low SNR for that redshift bin. }
    \label{fig:r0}
\end{figure}

\begin{figure}
    \centering
    \includegraphics[width=0.97\linewidth]{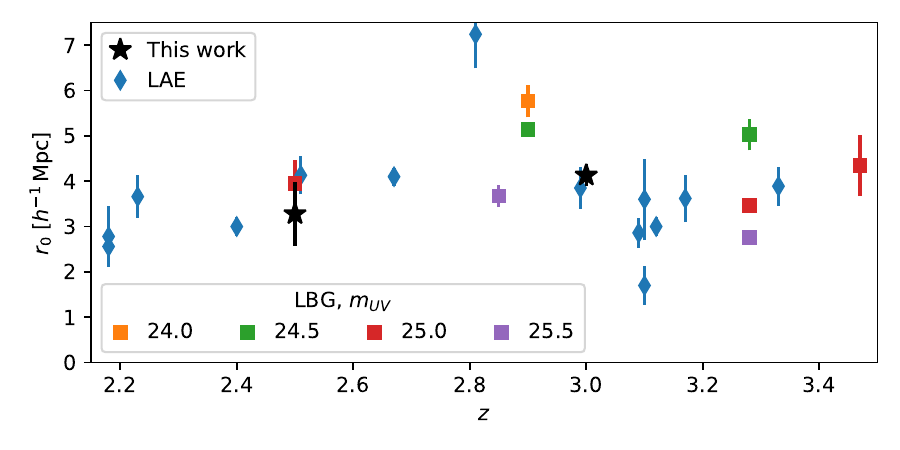}
    \caption{The best-fit $r_0$ inferred from our HOD-based mock fits to the data, compared to previous literature on LAEs \cite{Kovac07,Gawiser07,Ouchi10,Bielby16,Kusakabe18,Hao18,Hong19,Khostovan19,White24} and LBGs \cite{Hildebrandt09,Bian13,Bielby11,Durkalec15}. We only plot the $r_0$ of the `total' sample here, for visualization.
    }
    \label{fig:r0_compare}
\end{figure}

Figure \ref{fig:r0} shows the $\chi^2$ for each of the HOD models in our grid, plotted against the $r_0$ for that model (as determined from $\xi(r)$ in the N-body simulation).  The absolute value of the $\chi^2$ determines whether the model is statistically consistent with the data, with $\chi^2\simeq 14$ being $2\,\sigma$ exclusion for 7 degrees of freedom (d.o.f.). Further, we see (most clearly in the total sample at $z=3$) that the data follow an approximately parabolic shape with a moderate amount of scatter.  This illustrates the extent to which $r_0$ `predicts' the $\chi^2$ value and hence the degree to which there is additional information beyond this amplitude in the data for the model class described by our set of HODs.  That the $z=2.5$ points show no clear minimum in the $\chi^2$ vs.\ $r_0$ is a combination of limited numerical resolution (preventing us from examining lower $r_0$) and low SNR for that redshift bin. The $r_0$ of the best-fit model inferred for each redshift (as shown in Table \ref{tab:HOD_params}) is consistent with that from an array of past work \cite{Kovac07,Gawiser07,Ouchi10,Bielby16,Kusakabe18,Hao18,Hong19,Khostovan19,White24,Hildebrandt09,Bian13,Bielby11,Durkalec15}, as shown in Figure \ref{fig:r0_compare}. Note that each $r_0$ estimate varies in its definition of galaxy samples, including this work, contributing to the large scatter among data points. Finally, the best-fit model of the wide sample displays clustering that is slightly ($<10\%$) stronger than that of the total sample, a result that is expected given its brighter luminosity. However, this difference falls well within the uncertainties of our analysis and is therefore inadequate to derive a definitive conclusion.

\section{Inferences from simulations}

In this section, we will use the (standard) HODs from \S\ref{sec:HOD} that are consistent with the abundance and clustering of our samples to investigate the implications for galaxy-halo connection, scale-dependent bias and modeling.  We do not have the statistical precision to see significant differences in the clustering as a function of luminosity or redshift, though we do see differences in just the best-fitting models.  
At $z=2.5$ we barely detect clustering for the wide sample, although the mock halo masses that match the level of clustering in our data are consistent with past inferences of halo masses ($\sim 10^{10-11} M_\odot$ \cite{Ouchi20}) for LAEs.  At $z=3$ we have a stronger detection of clustering in both samples and our data prefer higher mass halos than the `typical' LAE in the literature, which are more consistent with traditional LBG halo masses $\sim 10^{12}\,h^{-1}M_\odot$ \cite{Giavalisco02,Shapley11,Wilson19}. 

\subsection{Mock Clustering}
\label{sec:mocks}

Our mock catalogs match the clustering and abundance of the samples selected in this work, and populate dark matter halos within a simulation based on the CDM paradigm.  To the extent that they provide reasonable proxies for the real galaxies it is useful to study their clustering properties.  Since we have access to much more information in the simulation than in observations, we calculate results in both real and redshift space and include cross-correlation with the dark matter field, to provide further insights into these high-$z$ galaxies which is of particular interest for their usefulness as cosmological tracers.

\begin{figure}
    \centering
    \includegraphics[width=0.97\linewidth]{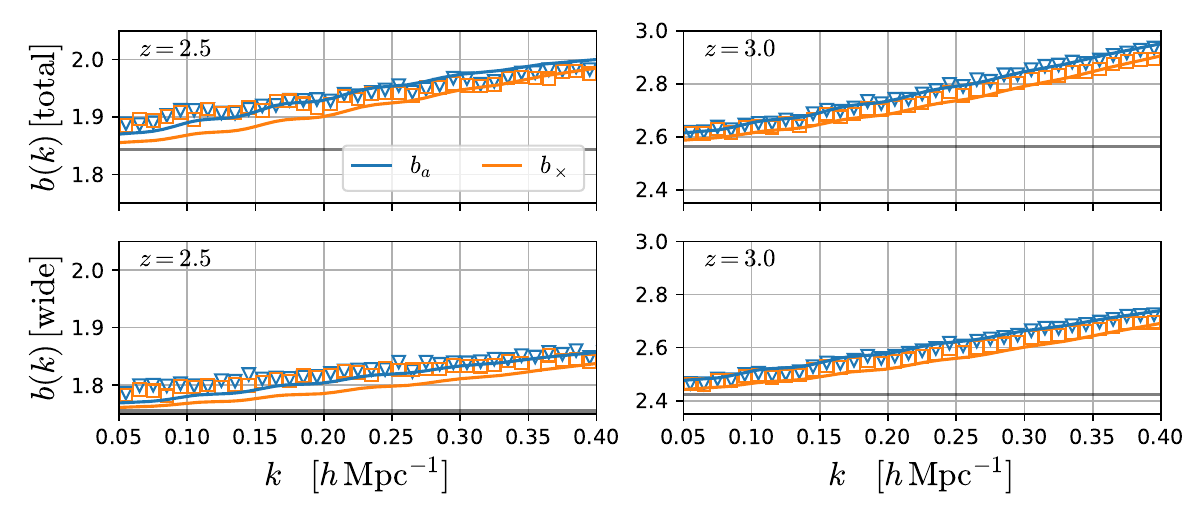}
    \caption{The real-space, scale-dependent bias in Fourier space for the mock clustering of two redshift bins (left: $z=2.5$, right: $z=3$) and two samples (top: total, bottom: wide).  In blue we show the auto-spectrum bias $b_a(k)=\sqrt{P_{gg}/P_{mm}}$ and in orange the cross-spectrum bias $b_\times(k)=P_{gm}/P_{mm}$, where we use the nonlinear matter spectrum for $P_{mm}$ and subtract Poisson shot noise from $P_{gg}$ to show maximum galaxy field-matter field correlation. 
    The solid lines are the PT predictions for the joint fit to the auto- and cross-spectrum in real and redshift space, where we see good agreement consistent within the limited statistics of the simulation. 
    At large scales, all biases asymptote to the linear bias $b_1$ (grey solid lines), as expected.
    }
    \label{fig:scalebias}
\end{figure}

For a cosmological tracer, the large-scale bias and its scale-dependence are key features important to cosmology. The results are shown in Figure \ref{fig:scalebias}, for both the auto-spectrum-derived bias $b_a=\sqrt{P_{gg}/P_{mm}}$ and cross-spectrum-derived bias $b_\times=P_{gm}/P_{mm}$, where $P_{gg}$, $P_{gm}$, and $P_{mm}$ are the real-space auto-, cross-, and nonlinear matter spectra and the ratio of these biases $b_\times/b_a$ is the correlation coefficient between the galaxy and matter fields. We subtract Poisson shot noise from the auto-spectra to show the maximum galaxy-matter correlation -- thus we would expect any observed sample to be less correlated with the matter field. We observe that both biases exhibit very modest scale dependence, with a $\sim 5\%$ evolution from large to small scales in the $z=2.5$ bin and $\sim15\%$ evolution in the $z=3$ bin. The fact that $b_a\approx b_\times$ for all samples indicates that galaxy and matter fields remain well correlated (in the absence of Poisson shot noise and RSD).  Given sufficient number density, these objects thus make excellent samples for probing large-scale structure and for cross-correlation studies.  Further notice how the analytical predictions discussed below (Table \ref{tab:biases}), shown in solid lines, agree well with the N-body predictions indicating that the samples can be accurately modeled.

\begin{table}[]
    \centering
    \begin{tabular}{c|c|c|c|c|c|c|c|c}
        $z$ & Sample &  $b_1$ &  $b_2$ &  $b_s$ &  $\alpha_0$ & $\alpha_2$ & $N_0$ & $N_2$ \\
        \hline
        \multirow{2}{1.5em}{2.5} & Total & $1.84$ & $1.02$ & $0.75$ & $-6.08$ & $2.11$ & $5.6\times 10^{1}$ & $-1.2\times 10^{2}$ \\
        \cline{2-9}
        & Wide & $1.76$ & $0.69$ & $0.50$ & $-3.24$ & $-0.64$ & $3.2\times 10^{1}$ & $2.5\times 10^{1}$ \\
        \hline
        \multirow{2}{1.5em}{3.0} & Total & $2.56$ & $-0.77$ & $-2.12$ & $-6.08$ & $5.99$ & $1.5\times 10^{2}$ & $-3.4\times 10^{2}$ \\
        \cline{2-9}
        & Wide & $2.42$ & $-0.40$ & $-2.78$ & $-5.19$ & $2.46$ & $8.9\times 10^{1}$ & $-1.1\times 10^{2}$ \\
    \end{tabular}
    \caption{The one-loop, Eulerian PT parameters for each clustering sample at $z=2.5$ and 3.0, in order of biases $b_i$, counterterms $\alpha_{2n}$, and stochastic terms $N_{2n}$. The parameters are inferred by a minimizer to the measurement of galaxy auto-spectrum in redshift-space, using ZCV \cite{Kokron22,DeRose23,Hadzhiyska23}, and galaxy-matter cross-spectrum in real-space. By using ZCV, the auto-spectrum can be measured precisely up to large scales.
    To improve the performance of the control variates technique, we do not randomly downsample the galaxies (to match $\bar{n}$) prior to measuring the clustering. Instead, we add Poisson shot noise to the original fit, using the number density from the target sample (Table \ref{tab:data}). The counterterms $\alpha_{2n}$ are in units $h^{-5}\text{Mpc}^5$ and stochastic terms $N_{2n}$ are in units $h^{-3-2n}\text{Mpc}^{3+2n}$.
    }
    \label{tab:biases}
\end{table}

In redshift space, we will model the mock clustering using the one-loop perturbative power spectrum, specifically with the publicly available software \texttt{velocileptors} \cite{Chen20a}. This modeling is identical to that adopted for the key results from two-point clustering in DESI DR1 \cite{DESI24-V}. The recent simulation-based study on high-redshift galaxies \cite{Sullivan25} also uses a variant of the theory, and it has been shown that the analysis results are consistent between variants of these perturbative models \cite{Maus24}.  In addition to the inclusion of non-linear dynamics and redshift space distortions the model has parameters for the non-linear bias relation, $\{b_i\}$, stochastic terms associated with shot noise and fingers of god, $\{N_{2n}\}$, and a set of ``counter terms'', $\{\alpha_{2n}\}$, that encapsulate small-scale physics not included in the model and that are dictated entirely by symmetry.

The biases, $\{b_i\}$, describe the galaxy overdensity field perturbatively 
\begin{equation}
    \delta_g \approx b_1 \delta + \frac{b_2}{2}\delta^2 + b_s s^2
\end{equation}
where we omit the third-degree bias $b_3$ as discussed in detail in refs.~\cite{Maus24,Maus24b}. The counterterms and stochastic terms each contribute to the power spectrum linearly
\begin{equation}
    P(k,\mu) \supset\alpha_{2n,a}\, \mu^{2n}(k/k_\star)^{2}P_{L}, \quad P(k,\mu) \supset N_{2n}(k\mu)^{2n}
\end{equation}
with $k_\star=1\,h\,\text{Mpc}^{-1}$. 
We refer the reader to refs.~\cite{VlaWhi19,Chen20a} for detailed discussion. 

\begin{figure}
    \centering
    \includegraphics[width=0.97\linewidth]{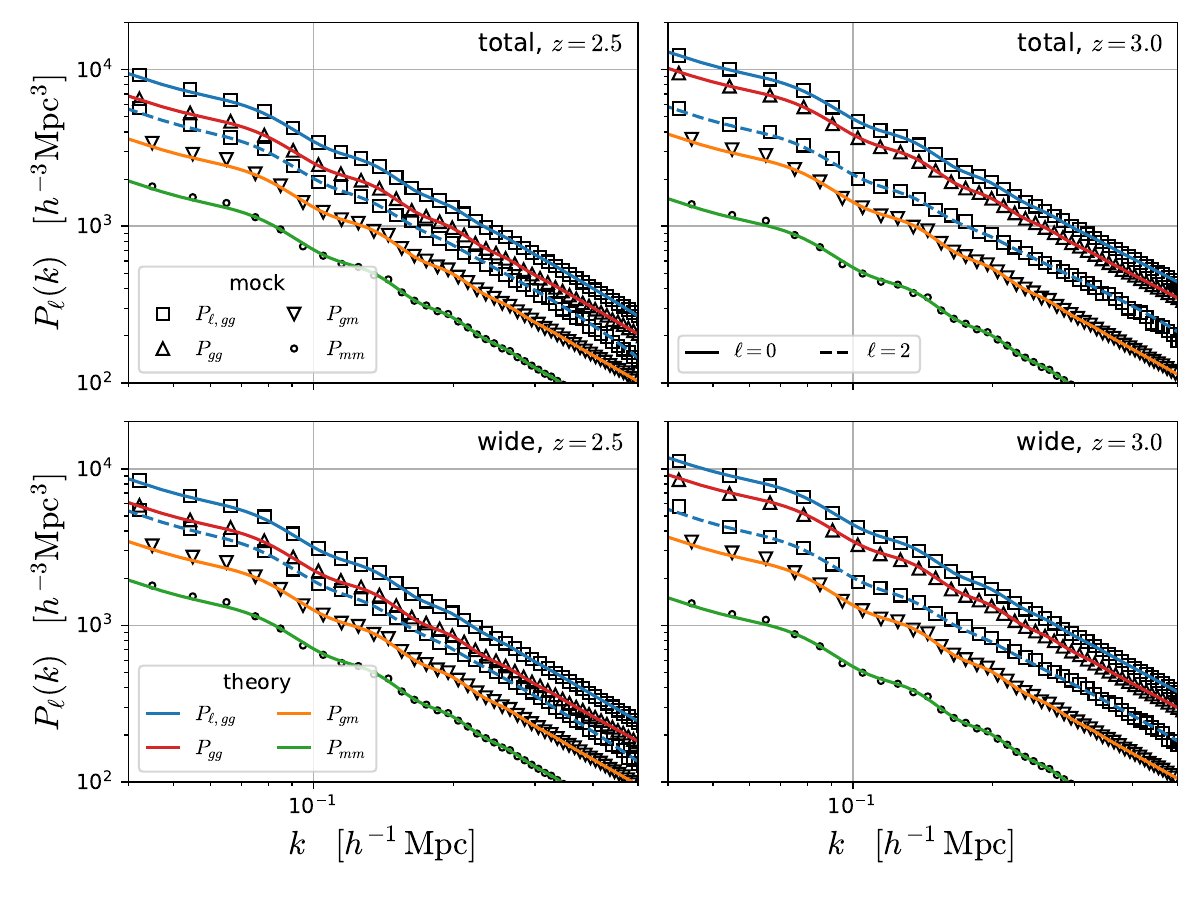}
    \caption{One-loop power spectrum fits to the mock catalog clustering, where the theory predictions are shown in solids lines and the mock measurements in markers. For each sample and redshift bin we simultaneously fit the redshift-space galaxy auto-spectrum (theory: blue, mock: squares) and the real-space galaxy auto- (red, upright triangle), galaxy-matter cross- (orange, reverse triange), and matter auto-spectrum (green, circle). The fitted parameters for each sample and $z$-bin are shown in Table \ref{tab:biases}.
    }
    \label{fig:PTfit1}
\end{figure}

One set of nuisance parameters is able to consistently model any clustering statistic of the galaxy, both in real- and redshift-space. We will take advantage of this to simultaneously model the multipoles ($\ell=0,\,2$) of the redshift-space galaxy auto-spectrum, $P_{\ell,gg}$, the real-space galaxy auto-spectrum, $P_{gg}$, galaxy-matter cross-spectrum, $P_{gm}$, and matter auto-spectrum, $P_{mm}$. 
In addition, although we are constrained to a $1\,h^{-1}{\rm Gpc}$ box, we are able to provide precise\footnote{An alternative means of overcoming sample variance is to do a field-level fit of the model to the mock density field \cite{Sullivan25}.  Our approach achieves similar performance, and has the agreeable side effect of generating a low-noise measure of $P_{\rm mock}$.} measurements of the mock power spectrum $P_{\rm mock}$ up to large scales using Zeldovich control variates \cite{Kokron22,DeRose23,Hadzhiyska23} and working at fixed cosmology. The bias fits at different redshifts, found through the minimizer implemented in the \texttt{Cobaya} code \cite{Cobaya,CobayaCode}, are shown in Figure \ref{fig:PTfit1}, with associated bias parameters shown in Table \ref{tab:biases}. We find that the linear biases $b_1$ at $z=3$ fall in the approximate range of $b_1$'s derived in the LAE simulations ($1.89<b_1<2.68$) of ref.~\cite{Sullivan25}, while falling significantly short of the $b_1$'s in LBG simulations ($3.92<b_1<4.24$). Note, however, that the disparity with the LBG simulations are not surprising, as the LBGs in this work have significantly fainter continuum than that modeled in ref.~\cite{Sullivan25} and LBG bias correlates strongly with continuum magnitude \cite{Wilson19}. 

The nonlinearity of the high-$z$ galaxy power spectra is a key quantity that we seek to characterize in these initial clustering analyses. The high-$z$ universe is known to host highly biased tracers, such as broadband-selected high-$z$ LBGs \cite{Wilson19,Ruhlmann-Kleider24}, and high bias usually implies strongly non-linear bias \cite{Desjacques18}. To what extent the medium-band galaxies in this study exhibit such nonlinearity will be important for future modeling efforts.

\begin{figure}
    \centering
    \includegraphics[width=0.97\linewidth]{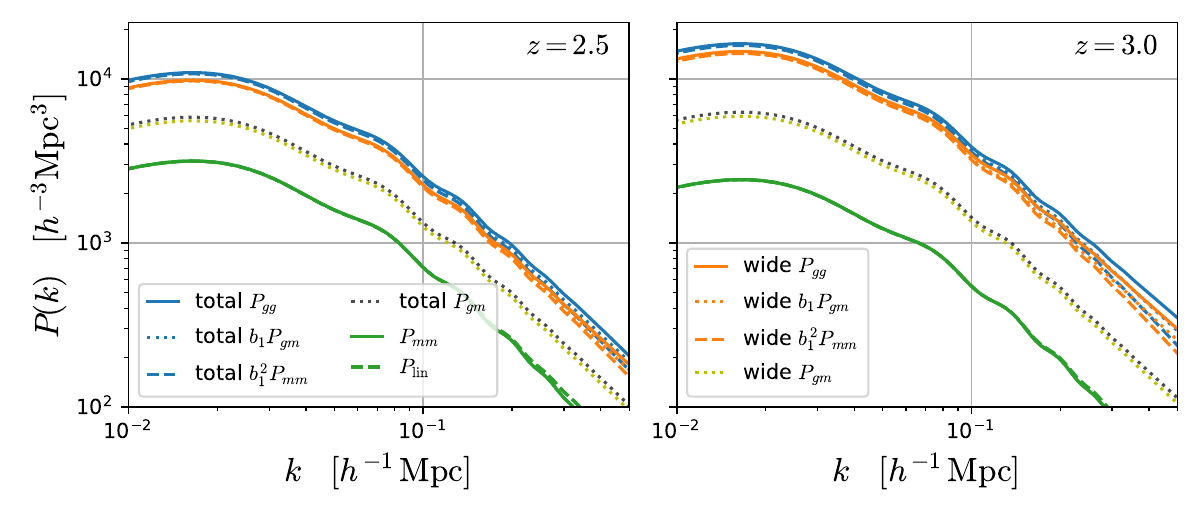}
    \caption{In blue and orange are comparisons between the real-space galaxy auto-spectrum, galaxy-matter cross-spectrum, and matter auto-spectrum (as computed using the best-fit PT model of Table \ref{tab:biases}; compared to N-body in Fig.~\ref{fig:PTfit1}).  In green is the comparison between the linear and non-linear matter auto-spectra, illustrating the nonlinearity of the matter field.  The blue and orange solid, dashed and dotted lines highlight the scale-dependence of the bias. The left and right panels correspond to the $z=2.5$ and $z=3$ results (Table \ref{tab:biases}), respectively (and the orange and blue lines almost overlap for $z=2.5$).
    }
    \label{fig:PTfit2}
\end{figure}

The nonlinearity of the power spectrum can be captured as a combination of two effects. One is the nonlinearity of the underlying density field. The decorrelation of the nonlinear matter field with the initial (Gaussian) density field reduces the information about the latter that can be learned from the former.  At higher redshifts the matter field remains linear to smaller scales (Appendix \ref{app:scales}). This can be observed from deviations between the linear matter spectrum $P_L$ and nonlinear spectrum $P_{mm}$.
The second effect is the nonlinearity due to scale-dependent bias, which indicates a decorrelation between the galaxy and matter density fields\footnote{The visually apparent difference between the spectra also rules out separable halo bias in the (analytical) halo model, which would imply that $P_{gm}/\sqrt{P_{gg} P_{mm}}$ is unity beyond the virial radius of halos. This agrees with recent work from ref.~\cite{Mons25}.   For further connections between PT and the halo model, see Appendix C of ref.~\cite{Maus24b}.}. This effect can be isolated by comparing $P_{gg}$, $b P_{gm}$, and $b^2 P_{mm}$, which would be identical in the absence of scale-dependent bias. We show both of these comparisons in Figure \ref{fig:PTfit2}, with the Poisson shotnoise subtracted for auto-spectra to emphasize scale dependence of clustering. While at $z=2.5$ both the non-linearity and scale-dependent bias are quite modest over the range of scales plotted, we see this is not true for the $z=3$ sample.  There the non-linearity in the matter field is smaller than at $z=2.5$, indicating the lower amplitude of large-scale structure growth, but the scale-dependent bias is much more pronounced.  For this sample, scale-dependent bias drives the decorrelation between the matter field and linear theory, requiring increased complexity (and more parameters) in the modeling. 

The FoG of these galaxies is another important modeling question that must be addressed. While there are no theoretical indications that the galaxy modeling of FoG will break down at high redshift, galaxies at high redshift tend to have higher bias and recent work using hydrodynamical simulations indicate that high-$z$ galaxies may exhibit FoG that range from smaller than that of low-$z$ galaxies \cite{Ravi24,Sullivan25} to stronger effects requiring higher-order perturbative terms \cite{Sullivan25}.
We observe that our PT models provide decent fits to the FoG of the best-fit models (Figure \ref{fig:PTfit1}) which is sufficient for our purposes. The mock clustering also shows that the zero-point crossing of the power spectrum quadrupole, which can be used as a proxy of FoG strength \cite{BaleatoLizancos25}, lies at $k>0.5\,h\,\mathrm{Mpc}^{-1}$, beyond our estimate of the non-linear scale, $k_{\rm nl}\approx\Sigma_{\rm Zel}^{-1}$ at $z\sim3$ (Appendix \ref{app:scales}; Fig.~\ref{fig:knl}). A small FoG term is also consistent with the measurements of line-of-sight correlations between LAEs and the Ly$\alpha$ forest measured in the spectra of LBGs \cite{Herrara25}.

\subsection{Forecast} \label{sec:forecast}

With the IBIS survey continuing to observe the sky, including deep coverage in the COSMOS field in 2025A (with spectroscopic follow-up with DESI) and wide fields in the future, we expect the observational constraining power to increase steadily in both 2D and 3D clustering. Hence, it is worthwhile considering what progress we may expect in the near future.

\begin{figure}
    \centering
    \includegraphics[width=0.98\linewidth]{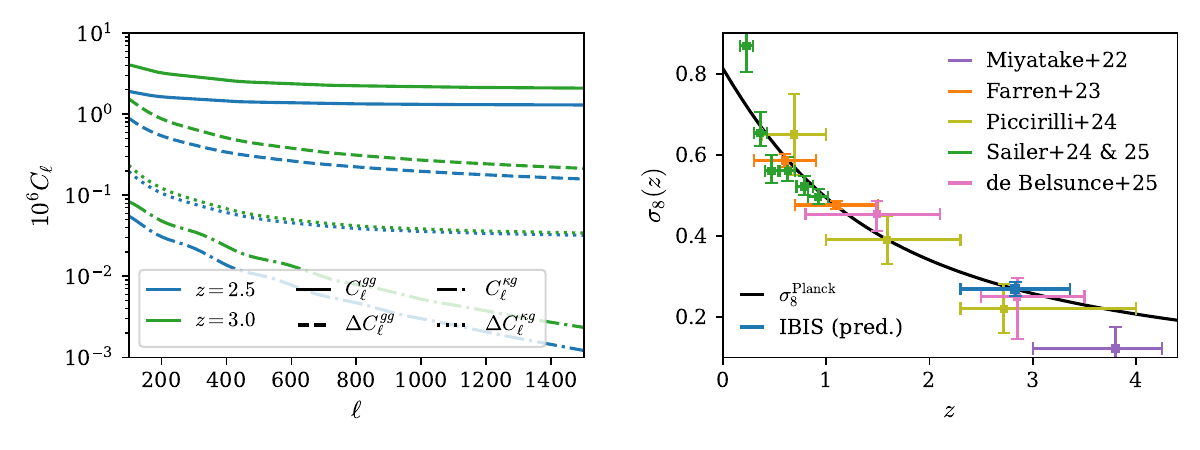}
    \caption{Left: Angular power spectra and their respective errors for the galaxy auto-spectrum and galaxy-lensing cross-spectrum for the third redshift bin (at $z\sim2.83$; corresponding to M$464$), assuming the galaxy characteristics of the wide sample and forecasts for SO lensing noise curves. The errors are calculated assuming a Gaussian covariance and $3000\deg^2$ sky area. Right: Fixed-shape $\sigma_8$ Fisher forecasts assuming the total sample over $3000\deg^2$ using \texttt{FishLSS} \cite{Sailer21}, in comparison with recent structure growth measurements obtained from galaxy-CMB lensing cross-correlations \cite{Sailer25,deBelsunce25,Sailer24,Farren24,Piccirilli24,Miyatake21}. The 6.5\% constraints on $\sigma_8(z)$ (blue) will be the tightest constraint on the clustering at such high redshift. This will improve to 5.1\% if the number density is doubled due to improved target selection.
    }
    \label{fig:cross-correlation}
\end{figure}

In the wide survey, measurement of angular clustering over a sky area of 3000$\deg^2$ will yield one of the most precise high-redshift clustering measurements to date. While the galaxy auto-correlation alone will show complete degeneracy between bias and clustering $\sigma_8$ (on linear scales), the introduction of galaxy-CMB lensing cross-correlation will effectively break this degeneracy \cite{Modi17,Wilson19,Miyatake21,Ebina24}. The expected power spectra and errors (for one redshift bin) are shown in the left panel of Figure \ref{fig:cross-correlation}, where the errors are calculated assuming Gaussian covariance. 

Using Fisher forecast software \texttt{FishLSS} \cite{Sailer21}, we are able to translate the observation expectations into projected constraints\footnote{For forecasting purposes, we use the $z$-range derived from the FWHM of the filter curves instead of that in Table \ref{tab:data} to avoid overlap between the $z$-bins, which is assumed when adding their respective Fisher matrices.}. Assuming the statistics of the wide subsample (which has the same luminosity limit as the wide survey) and `goal' sensitivity for Simons Observatory (SO) \cite{SimonsObs} lensing noise curves\footnote{The noise curves are generated by the \href{https://github.com/simonsobs/so_noise_models/blob/master/LAT_lensing_noise/lensing_v3_1_0/README.md}{SO Noise Calculator}.}, we find that IBIS will yield the tightest $\sigma_8$ constraint at high redshift after 3000$\deg^2$, as shown in the right panel of Figure \ref{fig:cross-correlation}. \footnote{If the footprint is altered from 3000$\deg^2$, constraints in Fisher forecasts will reduce by a factor of $\sqrt{f_{\rm sky, new}/f_{\rm sky, old}}$. This indicates that a footprint of 300$\deg^2$, which would degrade the constraints in Figure \ref{fig:cross-correlation} by a factor of $\sqrt{10}$, is sufficient to deliver the tightest constraints on $\sigma_8$ at $z\sim3$ at the Fisher forecast level.} We refer the reader to ref.~\cite{Sailer21} for a detailed description of fixed-shape $\sigma_8$ forecasts. 

We want to emphasize that the clustering analysis in this work deliberately refines the sample to have well-characterized $f_{\rm int}$ and $dN/dz$, at the expense of increased statistical noise. Due to this refinement, we can reliably measure $\sigma_8$ from cross-correlation, even without extensive spectroscopic follow-up.  However, if we can improve our selections, we can dramatically increase the constraining power of the sample. Undoing the refinement will quickly increase the sample size by a factor of $\sim2$ and ongoing work in target selection indicates that further improvement of another factor of $\sim2$ is likely.  We intend to report these investigations in a future publication.

\section{Extended models of HOD}
\label{app:HMQ}

\begin{figure}
    \begin{subfigure}{.43\textwidth}
      \centering
      \includegraphics[width=\linewidth]{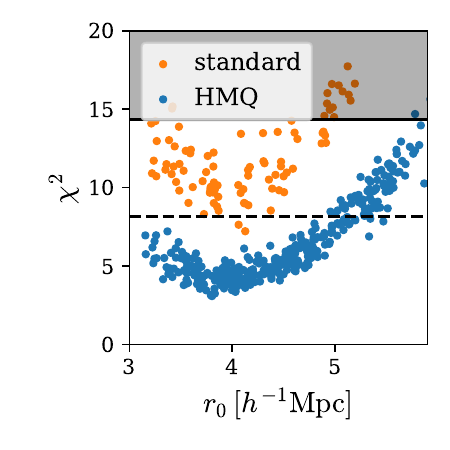}
    \end{subfigure}
    \begin{subfigure}{.57\textwidth}
      \centering
      \includegraphics[width=\linewidth]{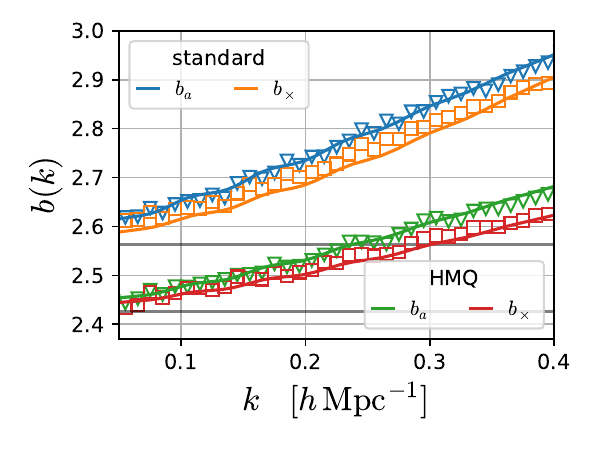}
    \end{subfigure}
    \caption{Left: The correlation length vs.\ goodness-of-fit for the `total' sample at $z=3$, using both the standard HOD (\S\ref{sec:HOD}) and HMQ HOD model.  The $1\,\sigma$ and $2\,\sigma$ CL for 7 d.o.f.\ are shown in dashed and solid lines, respectively, with the region above $2\,\sigma$ shaded in grey. We find that in both cases $r_0$ serves as an effective compressed statistic encoding the complexity of the HOD models, but the HMQ model is able to probe better fits, at lower $r_0$. Right: The scale-dependent bias of the standard and HMQ HOD models. Both the auto- ($b_a=\sqrt{P_{gg}/P_{mm}}$) and cross-bias ($b_\times=P_{gm}/P_{mm}$) are shown, with their ratio $b_a/b_\times$ showing the correlation between the galaxy and non-linear matter fields.}
    \label{fig:HMQ}
\end{figure}

It is far from clear that the ``standard'' HOD model (\S\ref{sec:HOD}) is appropriate for modeling our galaxy population.  In this section we consider an alternative to the standard HOD model explored above:  the High Mass Quenched (HMQ) HOD model. The HMQ model was initially proposed to model BOSS/eBOSS and DESI Emission Line Galaxies (ELGs) \cite{Alam20,Rocher23}. These ELGs are actively star forming galaxies with prominent O{\sc ii}.  As star formation is inefficient near the center of massive halos (at lower $z$) the HMQ model, which inhibits (very) massive halos from hosting central galaxies, were designed to better match our empirical understanding of ELG occupancy. The LAEs and LBGs that we consider in this work are also known to be actively star forming galaxies and hence merit analysis under this extended model. In addition, recent studies independently report results supporting this picture. Ref.~\cite{Sullivan25} has found that hydrodynamical simulations that fit to existing LAE data show a decline in central galaxy occupation past a halo mass of $\sim10^{11}\,h^{-1}M_\odot$, while ref.~\cite{Khoraminezhad25} has found that RT of Lyman alpha photons can significantly suppress the observation of central LAEs in massive halos.

In the HMQ model, the halo occupation of central and satellite galaxies are defined as
\begin{align}
\langle N_{\text{cen}}(M) \rangle &= 2A\phi(M)\Phi(\gamma M) + \frac{1}{2Q}\left[ 1+ \text{erf}\left(\frac{\log_{10}{M/M_{\rm cut}}}{0.01}   \right)  \right]  \\
\langle N_{\text{sat}}(M) \rangle &= \left( \frac{M - \kappa M_{\text{cut}}}{M_1} \right)^{\alpha}
\quad \text{for } M > \kappa M_{\text{cut}}
\end{align}
where $M$ is the halo mass, $\phi$ and $\Phi$ are normal distribution and cumulative distribution
\begin{equation}
    \phi(x) = \mathcal{N}(\log_{10}{M_{\rm cut}},\sigma_M) \quad , \quad 
    \Phi(x) = \int_{-\infty}^x \phi(t)dt = \frac{1}{2}\left[1+\text{erf}\left(\frac{x}{\sqrt{2}}\right) \right]
\end{equation}
while $A=[p_{\rm max}-1/Q]/\max[2\phi(x)\Phi(\gamma x)]$ and $\{\sigma_M$, $p_{\rm max}$, $\gamma$, $Q\}$ are the new model parameters, with $\sigma_M$ substituting $\sigma$.  The amplitude and width of the central HOD are controlled by $p_{\rm max}$ and $\sigma_M$ while $\gamma$ controls the low-$M$ turn-on and $Q$ the limit of $\langle N_{\rm cen}\rangle$ as $M\to\infty$. 

While a detailed comparison between different flavors of HOD models to investigate which models align better with high-$z$ galaxies is beyond the scope and precision of this work, we can compare the mocks made with the HMQ and `standard' HODs to gain some insights into how sensitive our inferences are to the HOD form.  For this we use the `total' $z=3$ sample, where the standard HOD results indicate that our simulation is most reliable, in terms of mass resolution.

The left panel of Figure \ref{fig:HMQ} shows $\chi^2$ vs.\ $r_0$ for a grid\footnote{We choose to vary between $\sigma_M=\{0.33,0.66\}$, $\alpha=\{0.33,0.66\}$, $Q=\{20,100\}$, $\gamma=\{1,5\}$, and $p_{\rm max}=\{0.33,0.66\}$, while sampling $\log_{10}M_{\rm cut}/(h^{-1}M_\odot)$ in steps of $0.25$ between $11.0$ and $12.0$ and $M_1/M_{\rm cut}$ in $\{10,30,100\}$.} of HMQ models, compared to the grid for the standard form discussed above.  We see the same general trends, with the HMQ models providing better fits than the standard HOD with less scatter; aligning with the results of refs.~\cite{Sullivan25,Khoraminezhad25}.  These results indicate that a single-parameter compression via $r_0$ is insufficient to capture the differences between the two HOD models.
In the right panel we compare the auto- and cross-spectrum biases for the best fits from the two grids of models.  The behavior is very similar, with approximately the same degree of scale-dependence and the same level of agreement that $b_a\approx b_\times$, though the best-fitting HMQ model has a slightly ($<10\%$) lower large-scale bias and scale-dependence than the standard form.  These differences are well within the error bars of our observations, and reflect (at least in part) the limitations of the data and our relatively coarse model grid.  To the extent that the large-scale bias, the scale dependence of the bias and the degree of decorrelation are among the principal factors impacting the use of tracers for cosmological inference it appears that our main conclusions are robust to changes in the functional form of the HOD.

\begin{figure}
    \centering
    \includegraphics[width=0.97\linewidth]{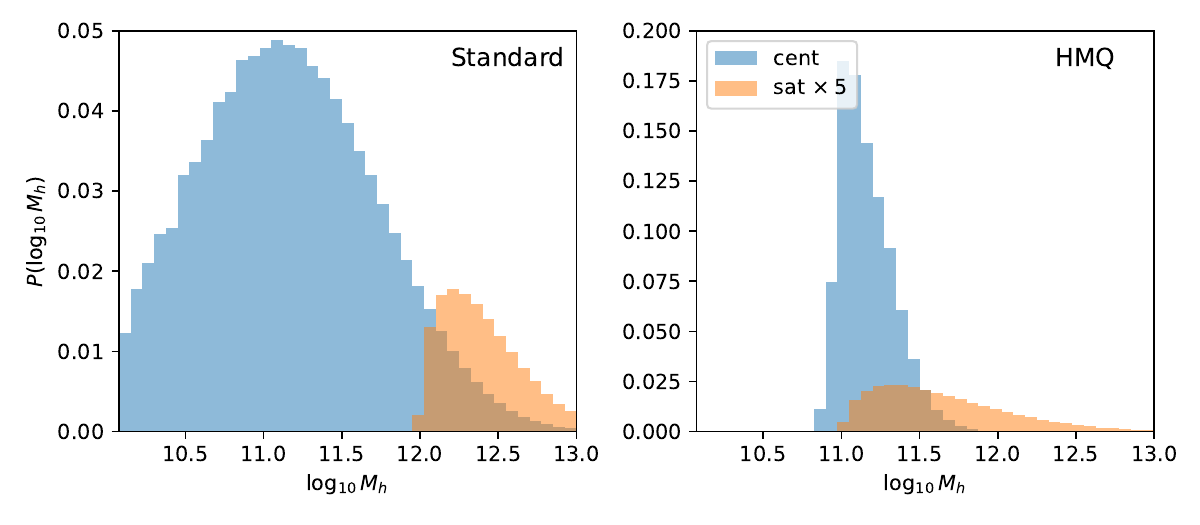}
    \caption{The host halo mass of central and satellite galaxies in the best-fit mock catalogs for the total sample at $z=3.0$, with the standard HOD considered in the main text (left) and the HMQ HOD considered in this appendix (right). The total (central $+$ satellite) distribution is normalized to integrate to unity and the satellite distribution is multiplied by 5 for visualization. The HMQ model predicts that the satellite galaxies are primarily hosted in low-mass halos, although with a significantly broad distribution extending to high-mass halos. 
    }
    \label{fig:Mh}
\end{figure}

Figure \ref{fig:Mh} shows the distribution of host halo masses for our mock galaxies for the two HOD forms.  In each case we have broken the distribution into that for the central and the satellite galaxies.  Modeling the galaxies within the standard form requires us to resolve significantly lower mass halos than for the HMQ form.  
The satellites in the HMQ model are less numerous ($f_{\rm sat}=0.06$) than the standard model ($f_{\rm sat}=0.10$, Table \ref{tab:HOD_params}) and reside in a much wider range of halo masses, suppressing the FoG effect.
Further comparison between HOD models, semi-analytic prescriptions and with hydrodynamic simulations will be valuable for future simulation based studies and mock catalog construction of high-$z$ galaxies \cite{Gurung-Lopez19,Ravi24,Sullivan25,Khoraminezhad25}.

\section{Conclusion} \label{sec:conclusion}

Over the last several decades, the study of large-scale structure has become an integral part of establishing and testing our cosmological model, with the latest generation of surveys pushing to ever greater statistical precision.  For example, the latest generation of BAO measurements has provided constraints on the distance scale with an aggregate precision well under a percent \cite{DESI-DR2}.  However the fraction of the available cosmological volume that these surveys have mapped is still small, and attention has been shifting to surveys of higher redshifts \cite{Schlegel22,Beseuner25}.  Such surveys promise transformational gains in cosmological constraining power and redshift lever-arm. However in order to properly plan for them we need to better understand the population of galaxies that would be used as targets.  In this paper we have begun such an investigation using observations from the IBIS and DESI surveys, with the goal of characterizing their clustering and halo occupations.  These, in turn, have direct implications for the design, data analysis and simulation requirements for future surveys. 

We have shown that the medium-band information from IBIS naturally allows us to define populations lying in well-delineated redshift shells of depth $\approx 200\,h^{-1}$Mpc spanning $z\simeq 2.3-3.4$.  These galaxies are typically faint in broad band magnitudes (e.g.\ median $r\simeq 25$, Figure \ref{fig:broadband}) but the presence of Ly$\alpha$ emission allows us to measure redshifts with $1-2\,$hr integrations on DESI \cite{Raichoor25}.  Our color cuts select objects that span the traditional divide betweeen LBGs and LAEs.  We estimate that despite the color cuts selecting for detectable Ly$\alpha$ peaks, almost half of the high-$z$ ($2.9<z<3.41$) are LBGs that would be selected by e.g.\ a $u$-dropout technique (\S\ref{sec:overlap}).

For our analysis, we have refined our sample to have a well-characterized interloper fraction, at the cost of increased shot noise. While this allows us to provide preliminary estimates of the clustering based on the angular data, for a spectroscopic survey we will likely choose to increase the target sample size at the expense of some uncertainty, as a dedicated spectroscopic survey can well characterize these uncertainties, at the expense of losing some observation efficiency. Further work on target selection optimized for a future spectroscopic campaign is ongoing.

The DESI observations for this program were designed to maximize the efficiency for obtaining redshifts while leaving fibers for other ancillary programs, and thus do not enable a direct 3D analysis (Appendix \ref{app:fiber}).  However, the natural decomposition of the sample into nearly disjoint shells of $\approx 200\,h^{-1}$Mpc depth allows us to make a detection of clustering through angular correlations (\S\ref{sec:angularclustering}, Appendix \ref{app:medium_shell}).  We find a marginal detection for one of the samples at $z=2.5$, though we are able to set interesting upper limits on the amplitude, and a convincing detection of clustering of the other sample at $z=2.5$, and both samples at $z=3$.  The angular clustering, along with $dN/dz$ and $f_{\rm int}$ from spectroscopic follow-up, then allows us to infer the real-space clustering. An introductory, model-agnostic analysis of these samples indicate correlation lengths of $r_0\sim 3-4\,h^{-1}$Mpc and linear bias $b\sim 1.8-2.5$, roughly in agreement with previous work on LAEs and LBGs. 

Further insight can be obtained by modeling our samples using HODs (\S\ref{sec:HOD}).  While the appropriate form of the HOD is not well known, we adopt the standard, five-parameter HOD model for the majority of our analyses in anticipation that it provides sufficient flexibility at the precision of this study.
Since our target selection is complex and galaxies with `strong' Ly$\alpha$ emission lines comprise only a fraction of the total galaxies at any epoch, we cannot simply use the abundance to determine the properties of the dark matter halos which host our targets.  Instead we use the clustering, in our case the projected correlation function on ``two halo scales''.  This provides us with a determination of the bias, which by assuming a particular functional form for the HOD, we can turn into an estimate of the characteristic halo mass.  We find that the halo masses are roughly consistent with previous estimates for LAE halo masses, with the exception of the wide sample at $z=3$ being closer to LBG halo mass estimates. 

The simulations that we employ had barely enough resolution to model the galaxy populations in this work.  This has direct implications for simulation requirements for future surveys.  Assuming the standard form of the HOD (Eq.~\ref{eqn:hod-cen}), to include the halos hosting the majority of the galaxies we need halos to be resolved down to $\approx 0.2\,M_{\rm cut}$, suggesting particle masses of two orders of magnitude lower than $M_{\rm cut}$.  Using the numbers for the $z=3$ sample, for example, we would need to resolve $\ge 10^{10}\,h^{-1}M_\odot$ halos (Fig.~\ref{fig:Mh}) implying a particle mass less than $\sim 10^9\,h^{-1}M_\odot$.  A simulation box subtending $30^\circ$ (for a survey area just under $1000\,\mathrm{deg}^2$) at $z=3$ would have a side length of $>2\,h^{-1}$Gpc (comoving) and require $\sim 10^{12}$ particles in order to resolve such halos.  Covering more area, or more cleanly resolving the host halo population, would increase the demands even further.  These requirements are relaxed by assuming a different form of the HOD (\S\ref{app:HMQ}), but remain very demanding even in this case.  This suggests that it will be necessary to carefully plan what simulations are really required for what tasks.

To the extent that our sample is characteristic of samples that will be utilized by further surveys, and that our mock catalogs approximate the true halo occupation of such samples, we can use clustering statistics measured from the mocks to inform us about the promise and modeling challenges ahead.  Our samples have a lower bias than dropout-selected LBGs at similar redshifts, have lower scale-dependent bias, are more correlated with the linear density field and have smaller fingers-of-god than those objects. 
In the long term these properties make our targets well suited for 3D clustering to constrain cosmology.  In the more immediate term we forecast that $3000\,\mathrm{deg}^2$ of IBIS data, in combination with CMB lensing, would allow a very strong measurement of the clustering amplitude of fluctuations at $z=2-4$. 

\section{Data Availability}

Software used for the analysis in this work are publicly available at \\ \href{https://github.com/HarukiEbina/IBIS-HOD}{\texttt{https://github.com/HarukiEbina/IBIS-HOD}}.

\section*{Acknowledgments}
HE and MW are supported by the DOE Office of Science under DE-SC0025523.
We thank Hendrik Hildebrandt for sharing the LBG galaxy catalog for cross-matching.
This work was performed in part at Aspen Center for Physics, which is supported by National Science Foundation grant PHY-2210452.
This material is based upon work supported by the U.S. Department of Energy (DOE), Office of Science, Office of High-Energy Physics, under Contract No. DE–AC02–05CH11231, and by the National Energy Research Scientific Computing Center, a DOE Office of Science User Facility under the same contract. Additional support for DESI was provided by the U.S. National Science Foundation (NSF), Division of Astronomical Sciences under Contract No. AST-0950945 to the NSF’s National Optical-Infrared Astronomy Research Laboratory; the Science and Technology Facilities Council of the United Kingdom; the Gordon and Betty Moore Foundation; the Heising-Simons Foundation; the French Alternative Energies and Atomic Energy Commission (CEA); the National Council of Humanities, Science and Technology of Mexico (CONAHCYT); the Ministry of Science, Innovation and Universities of Spain (MICIU/AEI/10.13039/501100011033), and by the DESI Member Institutions: \url{https://www.desi.lbl.gov/collaborating-institutions}. Any opinions, findings, and conclusions or recommendations expressed in this material are those of the author(s) and do not necessarily reflect the views of the U. S. National Science Foundation, the U. S. Department of Energy, or any of the listed funding agencies.
The authors are honored to be permitted to conduct scientific research on I'oligam Du'ag (Kitt Peak), a mountain with particular significance to the Tohono O’odham Nation.

This research draws upon DECam data as distributed by the Astro Data Archive at NSF NOIRLab. NOIRLab is managed by the Association of Universities for Research in Astronomy (AURA) under a cooperative agreement with the U.S. National Science Foundation.

This project used data obtained with the Dark Energy Camera (DECam), which was constructed by the Dark Energy Survey (DES) collaboration. Funding for the DES Projects has been provided by the US Department of Energy, the US National Science Foundation, the Ministry of Science and Education of Spain, the Science and Technology Facilities Council of the United Kingdom, the Higher Education Funding Council for England, the National Center for Supercomputing Applications at the University of Illinois at Urbana-Champaign, the Kavli Institute for Cosmological Physics at the University of Chicago, Center for Cosmology and Astro-Particle Physics at the Ohio State University, the Mitchell Institute for Fundamental Physics and Astronomy at Texas A\&M University, Financiadora de Estudos e Projetos, Fundação Carlos Chagas Filho de Amparo à Pesquisa do Estado do Rio de Janeiro, Conselho Nacional de Desenvolvimento Científico e Tecnológico and the Ministério da Ciência, Tecnologia e Inovação, the Deutsche Forschungsgemeinschaft and the Collaborating Institutions in the Dark Energy Survey.

The Collaborating Institutions are Argonne National Laboratory, the University of California at Santa Cruz, the University of Cambridge, Centro de Investigaciones Enérgeticas, Medioambientales y Tecnológicas–Madrid, the University of Chicago, University College London, the DES-Brazil Consortium, the University of Edinburgh, the Eidgenössische Technische Hochschule (ETH) Zürich, Fermi National Accelerator Laboratory, the University of Illinois at Urbana-Champaign, the Institut de Ciències de l’Espai (IEEC/CSIC), the Institut de Física d’Altes Energies, Lawrence Berkeley National Laboratory, the Ludwig-Maximilians Universität München and the associated Excellence Cluster Universe, the University of Michigan, NSF NOIRLab, the University of Nottingham, the Ohio State University, the OzDES Membership Consortium, the University of Pennsylvania, the University of Portsmouth, SLAC National Accelerator Laboratory, Stanford University, the University of Sussex, and Texas A\&M University.

Based on observations at NSF Cerro Tololo Inter-American Observatory, a program of NOIRLab (NOIRLab Prop. ID 2023B-184194; PI: A.~Dey and D.~Schlegel), which is managed by the Association of Universities for Research in Astronomy (AURA) under a cooperative agreement with the U.S. National Science Foundation.

\appendix

\section{Relevant scales}
\label{app:scales}

\begin{figure}
\begin{subfigure}{.49\textwidth}
  \centering
  \includegraphics[width=\linewidth]{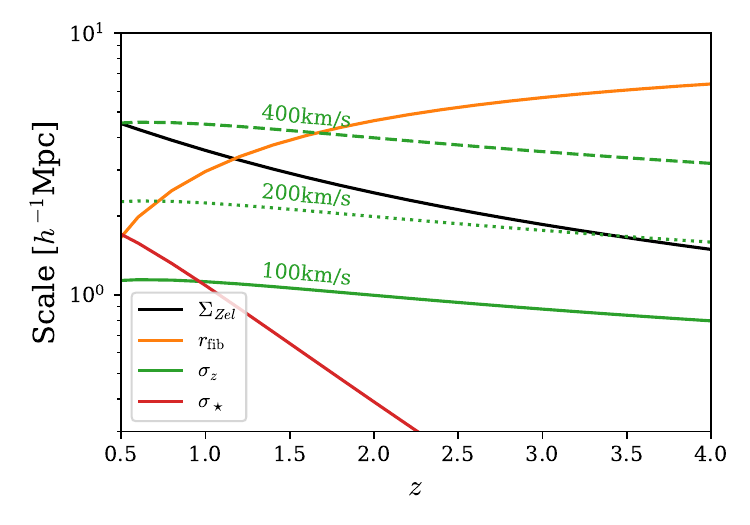}
\end{subfigure}
\begin{subfigure}{.49\textwidth}
  \centering
  \includegraphics[width=\linewidth]{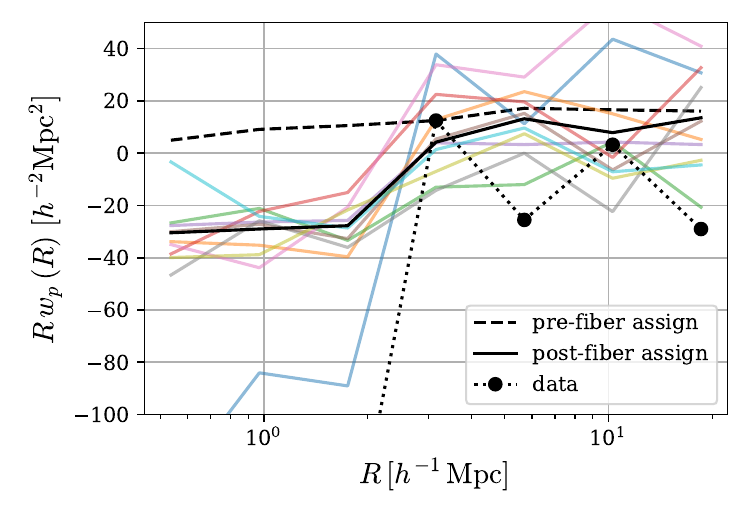}
\end{subfigure}
\caption{Left: Comparison of different `characteristic' scales as a function of redshift. The black solid line is an estimate of the non-linear scale, $\Sigma_{\rm Zel}$. The orange line is the transverse distance subtended by the DESI `fiber exclusion' scale, $0.05^\circ$. The green lines are the limitations on line-of-sight modes due to redshift error, with dashed, dotted, and solid corresponding to 400, 200, and $100\,\mathrm{km}\,\mathrm{s}^{-1}$ errors, respectively. The red line is the virial velocity of the characteristic halo mass. Right: Projected clustering measurement $w_p$ computed from a mock fiber assignment. We fiber assign using dithering, inducing a complex selection effect on the clustering and causing the post-assignment measurement (faint solid lines with mean shown as the black solid line) to deviate significantly from the pre-assignment measurement (mean shown as the black dashed line). The post-assignment clustering qualitatively reproduces the large-scale anti-correlation in the data (black circles) including the sign change at large $R$.
}
\label{fig:knl}
\end{figure}

In this appendix we briefly summarize some of the key physical scales entering into our calculations since at higher redshift different scales can dominate than at lower redshift where much of our intuition has been built.  We show these scales in Fig.~\ref{fig:knl}, as their evolution has important implications for the way in which survey design and survey non-ideality inform our analyses.

The non-linear scale can be estimated from the (comoving) Zeldovich displacement
\begin{equation}
    \Sigma_{\rm Zel}^2(z) = \int\frac{dk}{6\pi^2}\ P_{\rm lin}(k,z)
    \simeq \left[ 5.87\,D(z)\ h^{-1}\mathrm{Mpc} \right]^2
\end{equation}
where $D(z)$ is the linear growth rate, normalized to $D(z=0)=1$ and the numerical value is for the Planck cosmology \cite{PlanckLegacy18}.  We expect this scale to show the limits of perturbative models, and also to describe the characteristic scale at which the matter field decorrelates from the linear field (as described in Fig.~\ref{fig:PTfit2} the galaxy field can decorrelate at larger scales due to scale-dependent bias).

In a spatially flat Universe the comoving distance subtended by angle $\theta$ is simply $\chi\,\theta$ (with $\chi$ the comoving distance).  This scale is important for selection effects that operate in angle, such as fiber assignment in a multi-object, fiber-fed spectrograph.  Constraints on fiber placement depend upon several factors, but lead to an artificial suppression of pairs below a particular angular scale, which for DESI is around $0.05^\circ$ (e.g.~Fig.~2 of ref.~\cite{Pinon25}).  At $z>2$ the distance subtended exceeds $\Sigma_{\rm Zel}$, our estimate of the non-linear scale.

Expressing redshift uncertainties as $\Delta v/c=\Delta z/(1+z)$ and converting to comoving distance uncertainty assuming Hubble's law we have
\begin{equation}
    \sigma_z =  \frac{c\, \Delta z}{H(z)} = \frac{c(1+z)}{H(z)}\ \frac{\Delta v}{c}
\end{equation}
If the distribution of redshift errors were Gaussian, this would lead to a damping of power along the line of sight by $\exp[-k_\parallel^2\sigma_z^2]$.  Our galaxies are much fainter than a typical low-$z$ target and, in general, it is harder to provide precise redshifts for broad emission lines that are affected by radiative transfer and possible bulk flow of the emitting gas.  The characteristic redshift error is thus a larger contributor to the total error budget than is common for low-$z$ galaxies in spectroscopic surveys \cite{Wilson19,Sailer21}.  From repeat observations of galaxies and comparison to external data, DESI has estimated characteristic redshift errors of $\approx 200\,\mathrm{km}\,\mathrm{s}^{-1}$ \cite{Ruhlmann-Kleider24,Raichoor25}.  If this continues to hold, line-of-sight power damping due to redshift errors will remain subdominant to the impacts from non-linearity (scaling as $(1+f)\Sigma_{\rm Zel}\simeq 2\Sigma_{\rm Zel}$) at the redshifts of interest.

One of the largest sources of non-linearity in the redshift-space clustering of galaxies is stochastic terms, often referred to as fingers of god.  The size of this term is driven by the satellite fraction and the virial velocity of a halo that hosts on average one satellite galaxy \cite{Maus24b}, but we can get a feel for the scales involved by considering the virial velocity of an $M_\star$ halo (i.e.\ a halo of characteristic mass in the halo mass function).  A standard definition has $M_\star=(4\pi/3)\bar{\rho}_m R_\star^3$ where $R_\star$ is the comoving radius such that $\sigma_{\rm lin}(R_\star,z)=\delta_c\simeq 1.686$.  If we measure our halo masses such that $r_{\rm vir}$ encloses a mean density of $200\times$ the background density then $R_\star = (200)^{1/3}\, r_{\rm vir}$ for $M_\star$ halos.  The virial velocity associated with such a halo is $\sqrt{GM_\star/r_{\rm vir,phys}}=\sqrt{1+z}\sqrt{GM_\star/r_{\rm vir,com}}$, where we have distinguished `physical' and `comoving' radii.  Using the Friedman equation to rewrite the $(8\pi G/3)\bar{\rho}_m$ in the $GM_\star$ term and converting from velocity to distance units we have\footnote{If instead we measured our halo masses using the critical density definition the $\sqrt{\Omega_m(z)}$ would not appear.}
\begin{equation}
    \sigma_\star = \sqrt{\Omega_m(z)}\ 10\, r_{\rm vir,com}
                 = \frac{10\,\sqrt{\Omega_m(z)}}{(200)^{1/3}}\ R_\star
                 \simeq 1.71\ R_\star
\end{equation}
where the last step assumes the observations are at high enough $z$ that $\Omega_m(z)\simeq 1$.  By $z\simeq 3$ this characteristic scale has shifted to very low masses, and shifts very rapidly to even higher redshift due to the flatness of $k^3P_{\rm lin}$ at high $k$, so the inferred $\sigma_\star$ is quite low.

When considering $\sigma_\star$ we should bear in mind that our galaxies are quite biased and thus inhabit halos more massive than $M_\star$ \cite{Cole89}.  Indeed observations suggest that the most likely host halo mass does not evolve strongly with redshift \cite{Wechsler18}, unlike $M_\star$.  On the other hand, should the satellite fraction be low and the central galaxies moving with the center of mass of their host halo the virial dispersion is much reduced.  Thus $\sigma_\star$ should be taken as an `order of magnitude' estimate.  To the extent that our HOD fits reliably predict the halo occupancy of our sample a more refined estimate can be obtained from $\alpha_2$ and $N_2$.

Finally, we have omitted the virial radius of the halos hosting our galaxies as (a) it is a property of the samples themselves and not the underlying cosmology and (b) it is constant with redshift.  However a $10^{12}\,h^{-1}M_\odot$ halo has $r_{\rm vir,com}\simeq 200\,h^{-1}$kpc which suggests that the 1-halo term \cite{Wechsler18} remains very far below the non-linear scale and well below the reach of any of the clustering measurements described in this paper.  The Lagrangian radius of such halos (at $200^{1/3}r_{\rm vir,com}$) is however more comparable to the non-linear scale ($\Sigma_{\rm Zel}$) at high $z$, indicating that the complexities of scale-dependent bias are expected to become more important than gravitational non-linearity in the early Universe \cite{Modi:2019hnu,Wilson19}.

\section{Fiber assignment} \label{app:fiber}

The DESI spectroscopy provides redshifts for a fraction of the galaxy sample, in principle allowing us to perform a 3D clustering analysis (in redshift space). However, the effects of fiber assignment make this difficult. Although fiber assignment complications are not new to modern spectroscopy experiments, the situation here is unique due to two distinct reasons. The first is the high redshift of our sample. As fiber assignment, being a property of the focal plane, excludes pairs in angular coordinates, moving to higher redshift directly implies that the physical scale for fiber assignment subtends a larger transverse distance. When combined with the smaller non-linear scales at high redshift, fiber assignment may potentially be a much more difficult problem at high redshift than at low redshift. The second is the particular fiber assignment scheme adopted for the spectroscopic follow-up in this study (\S\ref{sec:DESI}). The fiber assignment was optimized to maximize the limited observation time, rather than being tailored to clustering studies, and in the process introduces density-dependent selections. 

DESI spectra are collected by assigning fibers to individual targets on the sky and the fibers are inherently limited in their proximity by the physical constraints of the focal plane.  In any single observation we cannot place fibers on two objects within $\sim 0.05^\circ$ of each other.  At low redshifts, this translates to relatively small scales ($r\lesssim1\,h^{-1}{\rm Mpc}$), whereas at $z\sim3$, it expands to around $r\sim 4 \, h^{-1}{\rm Mpc}$, shifting the information lost due to fiber assignment from halo/sub-halo scales to larger, perturbative scales. This redshift-dependent change in scale is illustrated in Figure \ref{fig:knl}. 

Such effects are not uncommon and many techniques have been proposed to deal with them (see e.g.\ \S 4 of \cite{Reid14}). 
One common method is to use weights for each object (sometimes referred to as independent inverse probability -- IIP -- weights) or each pair of objects (pairwise inverse probability -- PIP -- weights).  A typical weight could e.g.\ be calculated from the frequency of observation across many iterations of a stochastic fiber assignent process \cite{Bianchi17}. An alternative, data-driven approach simply assigns redshifts of unobserved galaxies to that of their closest neighbor with a successful redshift measurement or increments the weight of that object in the clustering \cite{Anderson12}.  Other methods upweight close pairs based on the angular clustering \cite{Hawkins03} or use alternative clustering statistics that exclude pairs that are close in angle \cite{Reid14,Pinon25}. 

Another approach is to give up on the full 3D information, and attempt to measure the projected correlation function, $w_p(R) =  \int dr_\parallel\, \xi(R,r_\parallel)$, via cross-correlation between the target (imaging) sample and spectroscopic sample. By using the target sample for one member of the pair, one is immune to pairwise spectroscopic incompleteness.  The projected correlation function can be expressed as a weighted sum over angular cross-correlation measurements \cite{Padmanabhan09,Myers09}.  However, a critical assumption in this computation is that the subsample of objects for which redshifts were obtained `fairly' samples the full distribution.  We will  see that in our case this assumption is invalid.

\begin{figure}
    \centering
    \includegraphics[width=0.97\linewidth]{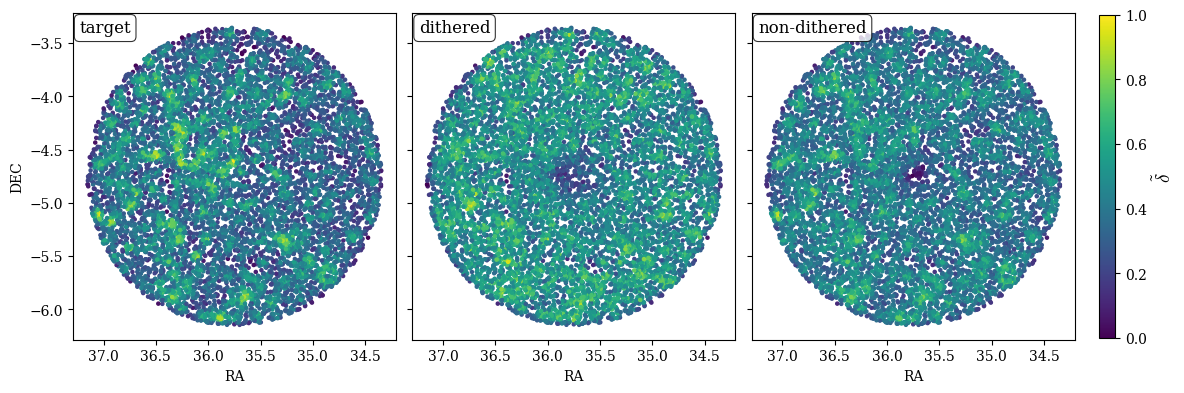}
    \caption{The impact of fiber assignment.  We show, using a mock catalog, an estimate of the projected density, $\tilde{\delta}$, across the field for: the original target sample (left), objects assigned a fiber with dithered fiber assignment (this work, center), and with non-dithered fiber assignment (typical, right).  The projected density estimate, $\tilde{\delta}$, is defined by the number of galaxies of the given type in a circle of radius $0.05^\circ$ (the fiber collision scale), normalized to run from 0 to 1.  Note that the dithered fiber assignment scheme produces notably smoother $\tilde{\delta}$ distributions than inherent in the target sample.
    }
    \label{fig:fiber_density}
\end{figure}

All of these methods to mitigate fiber assignment effects are ineffective for the samples at hand due to the unique fiber assignment scheme that was adopted for these observations.  In particular, the dependence of the final fiber assignment on the first round of fiber assignment introduces a density-dependent sampling, as shown in Figure \ref{fig:fiber_density} where we show the distribution of the local density measure $\tilde{\delta}$ defined by the normalized count of mock galaxies within a fiber assignment radius.

These difficulties even appear when attempting to correct for fiber assignment using the spectroscopic-photometric cross-correlation introduced above. We run the dithering fiber assignment over a mock catalog\footnote{Note that this mock isn't one of the best-fit catalogs introduced in the main text, as the fiber assignment effect must be modeled over the observed catalog, not the polished clustering catalog in this work.} and compare the pre- and post-fiber assignment $w_p$ from cross-correlation. 
As shown in the right panel of Figure \ref{fig:knl}, the data show not only ``small-scale'' pairwise incompleteness up to $R\sim 4h^{-1}\text{Mpc}$, but also significant anti-correlation at large scales. Both effects are unphysical, and the latter, in particular, cannot be corrected by any method that upweights close pairs. We can qualitatively reproduce these effects using mock catalogs, although we do not expect precise agreement as the complex sampling effect likely introduces dependencies on aspects of the mocks that are not tuned to be correct (e.g.\ higher order moments of the clustering or close-pair statistics like counts in cells). 
Given these results, we deem that a correct modeling of this complex sampling at high redshift requires far more precise and accurate mocks than we are able to produce at present.  Instead we use the angular clustering within the medium-band-selected shells (Appendix \ref{app:medium_shell}), whose interpretation depends only upon knowing $dN/dz$ and $f_{\rm int}$ and not on an understanding of the effects of fiber assignment.

\section{Angular clustering in medium-band selected samples}
\label{app:medium_shell}

\begin{figure}
    \begin{subfigure}{.49\textwidth}
      \centering
      \includegraphics[width=\linewidth]{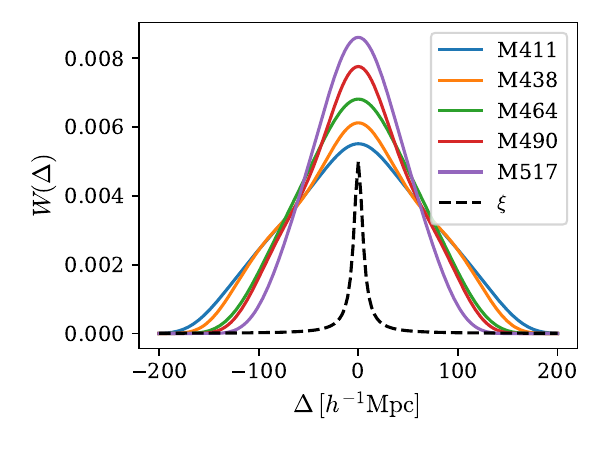}
    \end{subfigure}
    \begin{subfigure}{.49\textwidth}
      \centering
      \includegraphics[width=\linewidth]{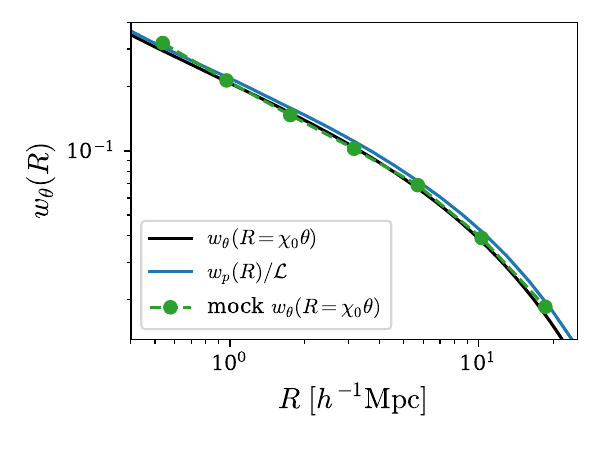}
    \end{subfigure}
    \caption{(Left) The line-of-sight window function, $W(\Delta)$, for the samples selected in our 5 filters.  For comparison, the dashed black line shows $\xi(r)\propto (R^2+\Delta^2)^{-1.8/2}$ as a function of $\Delta$ for $R=5\,h^{-1}$Mpc and arbitrary normalization. (Right) Comparison between our medium-band estimator, $w_\theta(R)$, averaged over our mock fields and the projected correlation function, $w_p(R)/\mathcal{L}$, computed from the full periodic $1h^{-1}\text{Gpc}$ box. The HOD model used is the best-fit model for the total sample at $z=3$.  The solid black line is the result of Eq.~(\ref{eqn:wtheta}) using $\xi$ measured from the periodic box.  The green dots show the average result of the mock pipeline for the M490 filter (with error bars suppressed), while the blue line then shows the approximation of Eq.~(\ref{eqn:wpR}).  With respect to our errors, the estimator is comfortably consistent with the true $w_p(R)$.  Since we use this approximation only to motivate our data combination, and not to actually perform fits to the data, we believe the degree of agreement is totally adequate.}
    \label{fig:los-window}
\end{figure}

Recall that if $f(\chi)$ defines the (normalized) redshift distribution of an individual shell, then the auto-correlation can be written as
\begin{equation}
    w(\theta) = \int d\chi_1 d\chi_2\ f(\chi_1) f(\chi_2)\, \xi\left(\sqrt{\Delta^2+\chi_1\chi_2\widetilde{\omega}} \right) 
\label{eqn:wtheta}
\end{equation}
where $\Delta=\chi_2-\chi_1$, $\widetilde{\omega}=2\sin(\theta/2)$, we have assumed that the correlation function does not evolve over the shell and that projection effects will `wash out' redshift-space distortions.  We can approximate $\chi_1\chi_2\simeq\bar{\chi}^2$, with $\bar{\chi}\equiv(1/2)(\chi_1+\chi_2)$, up to corrections of order $(\Delta/\bar{\chi})^2$.  Since $\Delta<0.1\,\bar{\chi}$ in our case this approximation introduces a sub-percent error.

The Jacobian for the change of variables $(\chi_1,\chi_2)\to (\bar{\chi},\Delta)$ is unity, so we immediately obtain
\begin{align}
  w_\theta(R)
  &\simeq \int d\bar{\chi}\,d\Delta\ f(\bar{\chi}-\Delta/2)f(\bar{\chi}+\Delta/2)\,\xi(\bar{\chi}\,\widetilde{\omega},\Delta) \\
  &\approx \int d\bar{\chi}\,d\Delta\ f(\bar{\chi}-\Delta/2)f(\bar{\chi}+\Delta/2)\,\xi(\chi_0\theta,\Delta) \\
  &= \int_{-\infty}^{\infty} d\Delta\ W(\Delta)\,\xi(R,\Delta)
\label{eqn:wR_xi}
\end{align}
where in the second line we have converted angles to (transverse) distances with $R=\chi_0\theta$ rather than $\bar{\chi}\,\widetilde{\omega}$ and we have defined
\begin{equation}
  W(\Delta) =  \int_0^\infty d\bar{\chi}\ f(\bar{\chi}-\Delta/2)f(\bar{\chi}+\Delta/2) \qquad .
\end{equation}
In the above we have used the small-angle approximation to clarify that the arguments of $\xi$ are $r_\perp$ and $r_\parallel$ with the integral over $r_\parallel$.  Corrections to the small-angle approximation are expected to be $\mathcal{O}(\theta^2)$, and are entirely negligible in our case.  Not negligible is the replacement $\bar{\chi}\to\chi_0$, i.e.\ that we can convert angles to transverse distances using the shell center ($\chi_0$) rather than $\bar{\chi}$.  This approximation is subdominant to our errors for our situation, where we are primarily determining how to combine disjoint slices, but introduces an error of $\mathcal{O}(\Delta\chi/\chi)$ that also depends upon the slope of $\xi$.

The line-of-sight window function, $W(\Delta)$, is the convolution of the radial selection function with itself.  For a tophat $f$ this would be a triangle function of twice the width (but the same overall area).  We show examples of $W(\Delta)$ for our slices in Fig.~\ref{fig:los-window}.  Note the width of $W$ is sufficiently large that we can approximate
\begin{equation}
    w_\theta(R) = \int_{-\infty}^{\infty} d\Delta
    \ W(\Delta)\,\xi(R,\Delta) 
    \approx W(0) \int d\Delta\ \xi(R,\Delta) 
    = \mathcal{L}^{-1} w_p(R)
\label{eqn:wpR}
\end{equation}
as in the main text.  For transverse scales much smaller than the characteristic scale of variation of $W$ the impact of redshift-space distortions is minimal.

Figure~\ref{fig:los-window}, right panel, shows how closely our mock $w_\theta(R)$ matches $\mathcal{L}^{-1}\,w_p(R)$ for the M490 sample.  The solid black line shows the result of Eq.~(\ref{eqn:wtheta}) using $\xi$ measured from the periodic box and can be seen to be in good agreement with the points that are estimated from the average of many mock observations.  The impact of the approximations made in going from Eq.~(\ref{eqn:wtheta}) to Eq.~(\ref{eqn:wpR}) can be seen in comparison to the blue line.  While the approximations are noticeably imperfect, they are well within the observational error bars (suppressed in Fig.~\ref{fig:los-window} but the same as in Fig.~\ref{fig:wt}).  It is important to note that in computing the HOD/theory predictions we follow the same procedure for the data and the mocks, so the relevant comparison is the green points to the black line.  The approximation in Eq.~(\ref{eqn:wpR}) is used only to motivate the manner in which we combine the samples adjacent in redshift (i.e.\ going from Fig.~\ref{fig:wt} to Fig.~\ref{fig:wR}). This may lead to non-optimal errors, but should introduce no bias.

\input{IBIS_author_affiliations_list}

\bibliographystyle{JHEP}
\bibliography{main}

\end{document}

%% file: IBIS_author_list.tex
\author[1,2,3]{{H.~Ebina}\orcidlink{0000-0002-1080-0955},}
\author[1,2,3]{{M.~White}\orcidlink{0000-0001-9912-5070},}
\author[3]{{A.~Raichoor}\orcidlink{0000-0001-5999-7923},}
\author[4]{{Arjun~Dey}\orcidlink{0000-0002-4928-4003},}
\author[3]{{D.~Schlegel},}
\author[5]{{D.~Lang},}
\author[6]{{Y.~Luo}\orcidlink{0000-0002-4623-0683},}
\author[3]{{J.~Aguilar},}
\author[7]{{S.~Ahlen}\orcidlink{0000-0001-6098-7247},}
\author[3]{{A.~Anand}\orcidlink{0000-0003-2923-1585},}
\author[8,9]{{D.~Bianchi}\orcidlink{0000-0001-9712-0006},}
\author[10]{{D.~Brooks},}
\author[11,12]{{F.~J.~Castander}\orcidlink{0000-0001-7316-4573},}
\author[3]{{T.~Claybaugh},}
\author[3]{{A.~Cuceu}\orcidlink{0000-0002-2169-0595},}
\author[13]{{K.~S.~Dawson}\orcidlink{0000-0002-0553-3805},}
\author[14]{{A.~de la Macorra}\orcidlink{0000-0002-1769-1640},}
\author[15,16]{{Biprateep~Dey}\orcidlink{0000-0002-5665-7912},}
\author[10]{{P.~Doel},}
\author[3,1]{{S.~Ferraro}\orcidlink{0000-0003-4992-7854},}
\author[17]{{A.~Font-Ribera}\orcidlink{0000-0002-3033-7312},}
\author[18,19]{{J.~E.~Forero-Romero}\orcidlink{0000-0002-2890-3725},}
\author[11,20,12]{{E.~Gaztañaga}\orcidlink{0000-0001-9632-0815},}
\author[3,21]{{S.~Gontcho A Gontcho}\orcidlink{0000-0003-3142-233X},}
\author[22]{{G.~Gutierrez},}
\author[23,24]{{H.~K.~Herrera-Alcantar}\orcidlink{0000-0002-9136-9609},}
\author[25]{{C.~Howlett}\orcidlink{0000-0002-1081-9410},}
\author[26]{{M.~Ishak}\orcidlink{0000-0002-6024-466X},}
\author[4]{{R.~Joyce}\orcidlink{0000-0003-0201-5241},}
\author[27]{{R.~Kehoe},}
\author[28]{{D.~Kirkby}\orcidlink{0000-0002-8828-5463},}
\author[3]{{T.~Kisner}\orcidlink{0000-0003-3510-7134},}
\author[3]{{A.~Kremin}\orcidlink{0000-0001-6356-7424},}
\author[10]{{O.~Lahav},}
\author[3]{{A.~Lambert},}
\author[3]{{M.~Landriau}\orcidlink{0000-0003-1838-8528},}
\author[29]{{L.~Le~Guillou}\orcidlink{0000-0001-7178-8868},}
\author[24]{{C.~Magneville},}
\author[30,17]{{M.~Manera}\orcidlink{0000-0003-4962-8934},}
\author[31,32,33]{{P.~Martini}\orcidlink{0000-0002-4279-4182},}
\author[4]{{A.~Meisner}\orcidlink{0000-0002-1125-7384},}
\author[34,17]{{R.~Miquel},}
\author[35]{{J.~Moustakas}\orcidlink{0000-0002-2733-4559},}
\author[36]{{E.~Mueller},}
\author[20]{{S.~Nadathur}\orcidlink{0000-0001-9070-3102},}
\author[24,3]{{N.~Palanque-Delabrouille}\orcidlink{0000-0003-3188-784X},}
\author[37,5,38]{{W.~J.~Percival}\orcidlink{0000-0002-0644-5727},}
\author[3,39,1]{{C.~Poppett},}
\author[40]{{F.~Prada}\orcidlink{0000-0001-7145-8674},}
\author[41]{{I.~P\'erez-R\`afols}\orcidlink{0000-0001-6979-0125},}
\author[42]{{G.~Rossi},}
\author[43]{{E.~Sanchez}\orcidlink{0000-0002-9646-8198},}
\author[44,45]{{M.~Schubnell},}
\author[3]{{J.~Silber}\orcidlink{0000-0002-3461-0320},}
\author[4]{{D.~Sprayberry},}
\author[45]{{G.~Tarl\'{e}}\orcidlink{0000-0003-1704-0781},}
\author[4]{{B.~A.~Weaver},}
\author[24]{{C.~Yèche}\orcidlink{0000-0001-5146-8533},}
\author[3]{{R.~Zhou}\orcidlink{0000-0001-5381-4372},}
\author[46]{{H.~Zou}\orcidlink{0000-0002-6684-3997},}

\affiliation{
\noindent \hangindent=.5cm $^{1}${University of California, Berkeley, 110 Sproul Hall \#5800 Berkeley, CA 94720, USA} \\
\noindent \hangindent=.5cm $^{2}${Department of Physics, University of California, Berkeley, 366 LeConte Hall MC 7300, Berkeley, CA 94720-7300, USA} \\
\noindent \hangindent=.5cm $^{3}${Lawrence Berkeley National Laboratory, 1 Cyclotron Road, Berkeley, CA 94720, USA} \\ \vspace{-2mm}

Rest of the affiliations are in Appendix \ref{sec:affiliations}.}

%% file: IBIS_author_affiliations_list.tex
\section{Author Affiliations}
\label{sec:affiliations}

\noindent \hangindent=.5cm $^{1}${University of California, Berkeley, 110 Sproul Hall \#5800 Berkeley, CA 94720, USA}

\noindent \hangindent=.5cm $^{2}${Department of Physics, University of California, Berkeley, 366 LeConte Hall MC 7300, Berkeley, CA 94720-7300, USA}

\noindent \hangindent=.5cm $^{3}${Lawrence Berkeley National Laboratory, 1 Cyclotron Road, Berkeley, CA 94720, USA}

\noindent \hangindent=.5cm $^{4}${NSF NOIRLab, 950 N. Cherry Ave., Tucson, AZ 85719, USA}

\noindent \hangindent=.5cm $^{5}${Perimeter Institute for Theoretical Physics, 31 Caroline St. North, Waterloo, ON N2L 2Y5, Canada}

\noindent \hangindent=.5cm $^{6}${Department of Physics \& Astronomy, University  of Wyoming, 1000 E. University, Dept.~3905, Laramie, WY 82071, USA}

\noindent \hangindent=.5cm $^{7}${Department of Physics, Boston University, 590 Commonwealth Avenue, Boston, MA 02215 USA}

\noindent \hangindent=.5cm $^{8}${Dipartimento di Fisica ``Aldo Pontremoli'', Universit\`a degli Studi di Milano, Via Celoria 16, I-20133 Milano, Italy}

\noindent \hangindent=.5cm $^{9}${INAF-Osservatorio Astronomico di Brera, Via Brera 28, 20122 Milano, Italy}

\noindent \hangindent=.5cm $^{10}${Department of Physics \& Astronomy, University College London, Gower Street, London, WC1E 6BT, UK}

\noindent \hangindent=.5cm $^{11}${Institut d'Estudis Espacials de Catalunya (IEEC), c/ Esteve Terradas 1, Edifici RDIT, Campus PMT-UPC, 08860 Castelldefels, Spain}

\noindent \hangindent=.5cm $^{12}${Institute of Space Sciences, ICE-CSIC, Campus UAB, Carrer de Can Magrans s/n, 08913 Bellaterra, Barcelona, Spain}

\noindent \hangindent=.5cm $^{13}${Department of Physics and Astronomy, The University of Utah, 115 South 1400 East, Salt Lake City, UT 84112, USA}

\noindent \hangindent=.5cm $^{14}${Instituto de F\'{\i}sica, Universidad Nacional Aut\'{o}noma de M\'{e}xico,  Circuito de la Investigaci\'{o}n Cient\'{\i}fica, Ciudad Universitaria, Cd. de M\'{e}xico  C.~P.~04510,  M\'{e}xico}

\noindent \hangindent=.5cm $^{15}${Department of Astronomy \& Astrophysics, University of Toronto, Toronto, ON M5S 3H4, Canada}

\noindent \hangindent=.5cm $^{16}${Department of Physics \& Astronomy and Pittsburgh Particle Physics, Astrophysics, and Cosmology Center (PITT PACC), University of Pittsburgh, 3941 O'Hara Street, Pittsburgh, PA 15260, USA}

\noindent \hangindent=.5cm $^{17}${Institut de F\'{i}sica d'Altes Energies (IFAE), The Barcelona Institute of Science and Technology, Edifici Cn, Campus UAB, 08193, Bellaterra (Barcelona), Spain}

\noindent \hangindent=.5cm $^{18}${Departamento de F\'isica, Universidad de los Andes, Cra. 1 No. 18A-10, Edificio Ip, CP 111711, Bogot\'a, Colombia}

\noindent \hangindent=.5cm $^{19}${Observatorio Astron\'omico, Universidad de los Andes, Cra. 1 No. 18A-10, Edificio H, CP 111711 Bogot\'a, Colombia}

\noindent \hangindent=.5cm $^{20}${Institute of Cosmology and Gravitation, University of Portsmouth, Dennis Sciama Building, Portsmouth, PO1 3FX, UK}

\noindent \hangindent=.5cm $^{21}${University of Virginia, Department of Astronomy, Charlottesville, VA 22904, USA}

\noindent \hangindent=.5cm $^{22}${Fermi National Accelerator Laboratory, PO Box 500, Batavia, IL 60510, USA}

\noindent \hangindent=.5cm $^{23}${Institut d'Astrophysique de Paris. 98 bis boulevard Arago. 75014 Paris, France}

\noindent \hangindent=.5cm $^{24}${IRFU, CEA, Universit\'{e} Paris-Saclay, F-91191 Gif-sur-Yvette, France}

\noindent \hangindent=.5cm $^{25}${School of Mathematics and Physics, University of Queensland, Brisbane, QLD 4072, Australia}

\noindent \hangindent=.5cm $^{26}${Department of Physics, The University of Texas at Dallas, 800 W. Campbell Rd., Richardson, TX 75080, USA}

\noindent \hangindent=.5cm $^{27}${Department of Physics, Southern Methodist University, 3215 Daniel Avenue, Dallas, TX 75275, USA}

\noindent \hangindent=.5cm $^{28}${Department of Physics and Astronomy, University of California, Irvine, 92697, USA}

\noindent \hangindent=.5cm $^{29}${Sorbonne Universit\'{e}, CNRS/IN2P3, Laboratoire de Physique Nucl\'{e}aire et de Hautes Energies (LPNHE), FR-75005 Paris, France}

\noindent \hangindent=.5cm $^{30}${Departament de F\'{i}sica, Serra H\'{u}nter, Universitat Aut\`{o}noma de Barcelona, 08193 Bellaterra (Barcelona), Spain}

\noindent \hangindent=.5cm $^{31}${Center for Cosmology and AstroParticle Physics, The Ohio State University, 191 West Woodruff Avenue, Columbus, OH 43210, USA}

\noindent \hangindent=.5cm $^{32}${Department of Astronomy, The Ohio State University, 4055 McPherson Laboratory, 140 W 18th Avenue, Columbus, OH 43210, USA}

\noindent \hangindent=.5cm $^{33}${The Ohio State University, Columbus, 43210 OH, USA}

\noindent \hangindent=.5cm $^{34}${Instituci\'{o} Catalana de Recerca i Estudis Avan\c{c}ats, Passeig de Llu\'{\i}s Companys, 23, 08010 Barcelona, Spain}

\noindent \hangindent=.5cm $^{35}${Department of Physics and Astronomy, Siena College, 515 Loudon Road, Loudonville, NY 12211, USA}

\noindent \hangindent=.5cm $^{36}${Department of Physics and Astronomy, University of Sussex, Brighton BN1 9QH, U.K}

\noindent \hangindent=.5cm $^{37}${Department of Physics and Astronomy, University of Waterloo, 200 University Ave W, Waterloo, ON N2L 3G1, Canada}

\noindent \hangindent=.5cm $^{38}${Waterloo Centre for Astrophysics, University of Waterloo, 200 University Ave W, Waterloo, ON N2L 3G1, Canada}

\noindent \hangindent=.5cm $^{39}${Space Sciences Laboratory, University of California, Berkeley, 7 Gauss Way, Berkeley, CA  94720, USA}

\noindent \hangindent=.5cm $^{40}${Instituto de Astrof\'{i}sica de Andaluc\'{i}a (CSIC), Glorieta de la Astronom\'{i}a, s/n, E-18008 Granada, Spain}

\noindent \hangindent=.5cm $^{41}${Departament de F\'isica, EEBE, Universitat Polit\`ecnica de Catalunya, c/Eduard Maristany 10, 08930 Barcelona, Spain}

\noindent \hangindent=.5cm $^{42}${Department of Physics and Astronomy, Sejong University, 209 Neungdong-ro, Gwangjin-gu, Seoul 05006, Republic of Korea}

\noindent \hangindent=.5cm $^{43}${CIEMAT, Avenida Complutense 40, E-28040 Madrid, Spain}

\noindent \hangindent=.5cm $^{44}${Department of Physics, University of Michigan, 450 Church Street, Ann Arbor, MI 48109, USA}

\noindent \hangindent=.5cm $^{45}${University of Michigan, 500 S. State Street, Ann Arbor, MI 48109, USA}

\noindent \hangindent=.5cm $^{46}${National Astronomical Observatories, Chinese Academy of Sciences, A20 Datun Road, Chaoyang District, Beijing, 100101, P.~R.~China}